\documentclass[twocolumn]{aastex631}

\usepackage{amsmath}

\begin{document}

\title{The Impact of Bars, Spirals and Bulge-Size on Gas-Phase Metallicity Gradients in MaNGA Galaxies}

\author[0000-0002-6505-9981]{M.E. Wisz}
\affiliation{Department of Physics, University of California, 5200 N Lake Rd, Merced, CA, 95343, USA}
\affiliation{Bryn Mawr College,
Department of Physics,
Bryn Mawr, PA 19010, USA}

\author[0000-0003-0846-9578]{Karen L. Masters}
\affiliation{Haverford College,
Departments of Physics \& Astronomy,
Haverford, PA 19041, USA}

\author[0000-0003-2594-8052]{Kathryne J. Daniel}
\affiliation{Department of Astronomy \& Steward Observatory, University of Arizona,
933 North Cherry Avenue,
Tucson, AZ 85721, USA}

\author[0000-0002-3746-2853]{David V. Stark}
\affiliation{Space Telescope Science Institute, 3700 San Martin Dr, Baltimore, MD 21218, USA}
\affiliation{William H. Miller III Department of Physics and Astronomy, Johns Hopkins University, Baltimore, MD 21218, USA}

\author[0000-0002-2545-5752]{Francesco Belfiore}
\affiliation{Istituto Nazionale di Astrofisica--Arcetri, Largo Enrico Fermi, 5, 50125 Firenze FI, Italy}

\begin{abstract}

As galaxies evolve over time, the orbits of their constituent stars are expected to change in size and shape, moving stars away from their birth radius. Radial gas flows are also expected. Spiral arms and bars in galaxies are predicted to help drive this radial relocation, which may be possible to trace observationally via a flattening of metallicity gradients. We use data from the Mapping Nearby Galaxies at Apache Point Observatory (MaNGA) survey, part of the fourth phase of the Sloan Digital Sky Surveys (SDSS-IV), to look for correlations of the steepness of gas-phase metallicity gradients with various galaxy morphological features (e.g. presence and pitch angle of spiral arms, presence of a large scale bar, bulge size). We select from MaNGA a sample of star forming galaxies for which gas phase metallicity trends can be measured, and use morphologies from Galaxy Zoo. We observe that at fixed galaxy mass (1)
the presence of spiral structure correlates with steeper
gas phase metallicity gradients; (2) spiral galaxies with
larger bulges have both higher gas-phase metallicities and
shallower gradients; (3) there is no observable difference with azimuthally averaged radial gradients between
barred and unbarred spirals and (4) there is no observable difference in gradient between tight and loosely wound spirals, but looser wound spirals have lower average gas-phase metallicity values at fixed mass. We discuss the possible implications of these observational results. 
\end{abstract}

\keywords{galaxies: abundances; ISM: abundances; galaxies: spiral; galaxies: evolution}

\section{Introduction} 
\label{sec:intro}
 Understanding the metal content of the interstellar medium (ISM) of a galaxy is a powerful technique to disentangle its cosmic evolution and the balance of cosmological gas inflow alongside the regulation of successive generations of star formation in a galaxy. In the current cosmological model, Big Bang nucleosynthesis creates a well understood mix of hydrogen, helium, and trace amounts of other atoms \citep{Steigman2007}. All other atoms (``metals") are created in galaxies through generations of stellar evolution enriching the ISM. 
 Meanwhile inflow of pristine gas from the cosmological environment of galaxies is generally understood to be required to maintain galaxy star formation rates to the current day, and create local disc galaxies \citep[see][for a relatively recent review]{SanchezAlmeida2014}. This accretion brings in metal-poor (or even metal-free) gas to the ISM in a galaxy, while galactic winds can generate outflows to remove metal rich gas from a galaxy \citep{Larson1974}. 
 
 In external galaxies, we can measure the ``metallicity" of a galaxy stellar population modeling techniques (e.g. \citealt{Conroy2009,Sanchez2016} which return the light- or mass-weighted average metallicity of a population of stars), or using emission lines to constrain the metallicity of the ionized gas in galaxy. Stellar metallicity encodes the metallicity of the ISM at the time of the star's birth, while gas-phase metallicity provides the current value. Gas-phase metallicity can be calculated using various different emission line calibrators, depending on the type of object and the emission lines available \citep[see][for a recent review]{Kewley2019}. 

 The advent of large scale spectroscopic surveys, such as the first phase of the Sloan Digital Sky Survey \citep[SDSS,][]{York2000}, made it possible to measure stellar and gas phase metallicities for large samples of external galaxies. 
 \citet{Tremonti2004} measured gas-phase oxygen abundances for nearly 53,000 SDSS galaxies, finding a tight correlation between stellar mass and metallicity, Z, (the MZR which shows that more massive galaxies are more metal enriched) which they attributed to the importance of galactic winds removing metal enhanced material effectively from lower mass galaxies. The Fundamental Metallicity Relation \citep[FMR,][]{Mannucci2010} recast the \citet{Tremonti2004} MZR, as one axis of a 3D mass-metallicity-star formation rate (SFR) relation, noting that especially for lower stellar mass galaxies ($\log(M_\star/M_\odot)<10.7$), metallicity is seen to decrease with increasing SFR. 

These results for the most part treat galaxies as single points and for large angular size galaxies, measure only values in the central regions which cover different physical fractions of galaxies at different redshifts. This is particularly undesirable since it has been known for some time that local disc galaxies, including our own Milky Way, show significant negative metallicity gradients (e.g. \citealt{Shaver1983,Ferguson1998,vanZee1998}). The current generation of integral field unit (IFU) surveys, such as the Calar Alto Legacy Integral Field Area Survey \citep[CALIFA, ][]{Sanchez2012, Sanchez2014}; Mapping Nearby Galaxies at Apache Point Observatory \citep[MaNGA,][]{Bundy2015}; and The Sydney-Australian-Astronomical-Observatory Multi-object Integral-Field Spectrograph Survey  \citep[SAMI,][]{Croom2012}, has opened up measurement of radial metallicity trends, in both stars and gas, to much larger samples of galaxies, and cements their use as an important diagnostic for galaxy evolution. Previous work measuring gas-phase radial metallicity trends of galaxies from data in these local Universe IFU surveys, has found that as an overall trend that gradients steepen with stellar mass \citep[e.g.][]{Perez-Montero2016,Belfiore2017}. There is also some evidence of large scale morphological dependence for gradients \citep[e.g.][]{SanchezMenguiano2016,Sanchez-Menguiano2018,Boardman2021,Barrera-Ballesteros2023,Pilyugin2024}, however local impacts may be less prominent \citep{Zinchenko2019}. Azimuthal variations in metallicity are also observed around the spiral arms in the Solar Neighborhood \citep{Hawkins23}. 

In this paper, we investigate if there is any observable impact of non-axisymmetric internal galaxy morphology (specifically spiral arms and bars) on radial gas-phase metallicity trends observed in a sample of 2,632 isolated star forming MaNGA galaxies.  This was motivated by simulations suggesting even the pitch angle of spiral arms might change the rate of radial relocation of stars \citep{Danielinprep}. Other work has also previously suggested morphological features might change gas-phase metallicity gradients. For example \citet{Orr2022} and others suggest spiral arms can act as "metal freeways" driving metal poor gas inwards, and metal rich gas outwards (for more on spiral arms see \citet[]{vilas-costas1992, Zaritsky1994}. Similar effects have been proposed for bars \citep[e.g.][]{martin1994,Athanassoula1992,Fragkoudi2016, kaplan2016,Zurita2021}.

The structure of the paper is as follows. In Section \ref{sec:data}, we introduce our use of the MaNGA and Galaxy Zoo (GZ) surveys. Section \ref{sec:sample selection} outlines the selection criteria that we follow to obtain our star-forming (SF) sample, or our \emph{SF parent sample}, in addition to all sub-samples and Section \ref{sec:methods} details how we obtain radial gas-phase metallicity gradients for our sample and generate population trend lines. In Section \ref{sec:results}, we present radial metallicity trends as well as various methods of accounting for the MZR in each of our sub-samples. Finally, Section \ref{sec:discussion} contains a discussion of our results. 

\section{Data and Sample Selection} \label{sec:data} 
In this work we make use of the final data release from the MaNGA survey \citep{Bundy2015} which was part of the 17th data release from SDSS \citep[][hereafter DR17]{DR17}. We also make use of GZ classification of MaNGA galaxies \citep[][]{Willett2013,Hart2016}, specifically the version released as a MaNGA Value Added Catalog with DR17. 

\subsection{MaNGA data}
\begin{figure*}
    \centering
    \includegraphics[width = 0.9\textwidth]{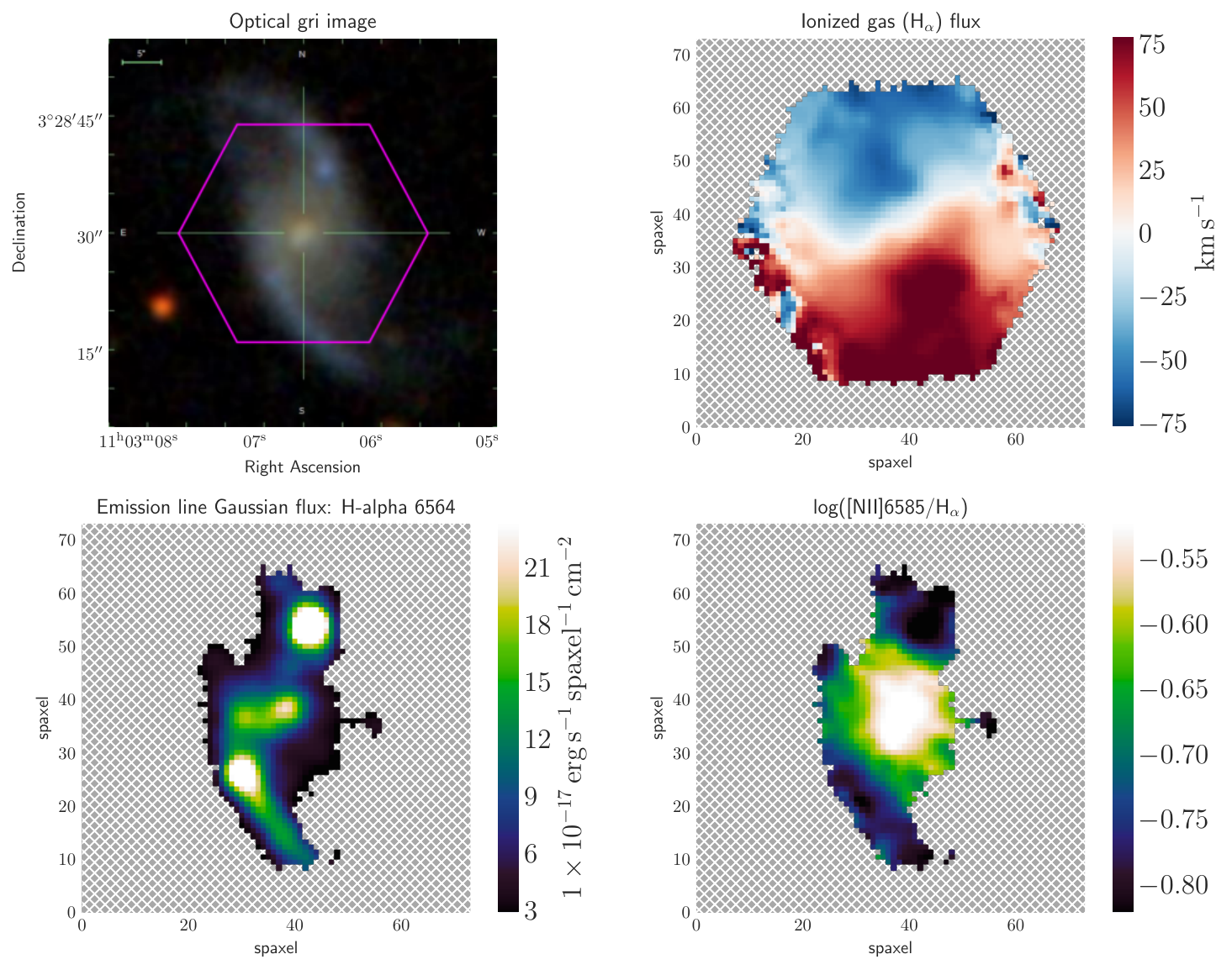}
    \caption{Example data for a spiral galaxy (MaNGA-ID: 1-62035) with loosely wound spiral arms evident in both the optical image (upper left) and traced out by the $H\alpha$ (lower left) and $[NII]$ flux maps (not shown). \emph{Top left:} optical $gri$ image of galaxy. Purple hexagon overlay is MaNGA field of view. \emph{Top right:} Spaxel map of ionized gas velocity for $H\alpha$. \emph{Bottom left:} Masked spaxel map of $H\alpha$ flux. \emph{Bottom right:} Masked spaxel map of $[NII]/H\alpha$ flux.}
    \label{fig:examplegalaxyanddata}
\end{figure*}

MaNGA \citep{Bundy2015} was a sub-survey of the fourth phase of the SDSS \citep[SDSS-IV,][]{Blanton2017} that used IFUs, made up of hexagonally close packed bundles of fibers, to obtain spatially resolved spectra across the face of 10,000 nearby galaxies. The MaNGA survey used the MaNGA Integral Field Unit Fiber Feed System \citep{Drory2015} which was built on the BOSS spectrograph \citep{Smee2013}. The survey operated at the Apache Point Observatory, using the SDSS telescope \citep{Gunn2006}. The MaNGA instrument carries various fiber bundles serving different purposes, including 17 science IFU bundles and 12 flux calibration bundles. The 10,000 galaxies in MaNGA were selected out of the NASA Sloan Atlas, (NSA; \citealt{Blanton2011}) to have a flat number density across log($M_{\star}/M_\odot) = 9-11$, in order to create a representative sample of galaxies \citep{Wake2017}. Additionally, the galaxies were selected such that two-thirds of the sample was surveyed out to 1.5 effective radii ($R_e$; the radius enclosing 50\% of the light as measured in the NSA; \citealt{Blanton2011}) and one-third out to 2.5$R_e$. These subsets are known as the Primary and Secondary samples respectively. Other smaller subsets were also included for specific scientific reasons. For more specifics of the design of the MaNGA sample, including weights to reproduce a volume limited sample, we refer the readers to \citet{Wake2017}. 

MaNGA differs from previous SDSS surveys, which only obtained spectra at the center of the galaxies, by obtaining spectra across the faces of the galaxies well beyond their nuclei. 
An example image of a MaNGA galaxy and maps of data products generated using the {\tt Marvin} Python module \citep[built for interacting with MaNGA data;][]{Cherinka2019} is shown in Fig. \ref{fig:examplegalaxyanddata}, where the purple hexagon in the top left panel marks the MaNGA field of view over laid on the optical $gri$ image of the example galaxy. The top right panel shows the ionized gas velocity (the H$\alpha \lambda 6564$ flux), the bottom left panel displays the masked H$\alpha \lambda 6564$ gas flux and finally the bottom right panel shows the masked ${\rm [NII} \lambda 6585] / [H\alpha$] gas flux ratio (which as we discuss in Section \ref{sec:methods} is used in some gas phase metallicity measures). The mask applied to the [H$\alpha$] flux and the ${\rm [NII} \lambda 6585] / [H\alpha$] gas flux ratio panels is the same mask as described later in Sec. \ref{sec:sample selection} that we use to create our samples. The [H$\alpha$] flux panel shows that in this galaxy, this emission line clearly trace the spiral arms that are seen in the optical $gri$ image. This data in MaNGA maps is provided in a grid of spatial pixels, or ``spaxels," where each spaxel is 0.5" wide. This is a sampling of the spatial resolution created by the MaNGA observations which are created by dithering of 2" fibers in typical seeing of 1.5\arcsec; the reconstructed point-spread-function (resolution) is measured as having a median value of 2.54" \citep{Yan2016}.

\subsection{Galaxy Zoo for MaNGA Galaxies}
GZ \citep[e.g.][]{Lintott2008,Willett2013} is a citizen science project uses volunteer contributions to visually classify galaxies. GZ has run over several generations of imaging surveys. We use classifications from the 2nd phase of GZ, or GZ2 \citep{Willett2013}, which used of images from the SDSS \citep{York2000} of galaxies in the Main Galaxy Sample \citep{Strauss2002}.  Due to catalog matching and magnitude limit choices, not all MaNGA galaxies are found in the initially released version of GZ2 \citep{Willett2013}. However, the Galaxy Zoo project provided a curated list of classifications for 96\% of the MaNGA sample, analyzed using the same methods as GZ2 and which was released as a Value Added Catalog in SDSS DR17 \citep{DR17}\footnote{Available at \url{ https://www.sdss4.org/dr17/data_access/value-added-catalogs/?vac_id=galaxy-zoo-classifications-for-manga-galaxies}}; it is this GZ catalog we use here. Specifically we make use of debiased, weighted vote fractions (fraction of users indicating they saw a specific feature) for (1) $p_{\rm features}$, answering no to ``is this galaxy simply round and smooth, with no sign of a disk?"; (2) $p_{\rm notedgeon}$, answering no to ``could this be a disk viewed edge-on?"; (3) $p_{\rm spiral}$, a yes to ``is there any sign of a spiral arm pattern?"; (4) $p_{\rm merger}$, the fraction of people who indicated they saw a merging system; (4) $p_{\rm just noticeable}$, $p_{\rm obvious}$, and $p_{\rm dominant}$, three of the four answers to ``how prominent is the central bulge, compared with the rest of the galaxy?"; (5) $p_{\rm bar}$, a yes to ``is there a sign of a bar feature through the center of the galaxy"; and finally (6) $p_{\rm medium}$ and $p_{\rm tight}$, and implicitly, $p_{\rm loose} = 1 - p_{\rm medium} - p_{\rm tight}$, the possible answers to ``how tightly wound do the spiral arms appear?". These classifications have been extensively used in observational extragalactic work over the last decade, and demonstrated to be reliable and objective quantitative visual classifications \citep[e.g. see][]{Willett2013,Masters2011,Hart2017,Masters2019,Mengistu2023}.

\subsection{Sample Selection}
\label{sec:sample selection}
Our initial sample consists of a total of 9,585 galaxies that have MaNGA observational data, and data from GZ2 \citep{Willett2013}, Pipe3D \citep{Sanchez2022} and the GEMA \citep[Galaxy Environment for MaNGA, ][]{Argudo2015} value-added catalogs. From this sample, we select a sample of galaxies that meet the following criteria:
\begin{figure}
    \centering
    \includegraphics[width = 3.2 in]{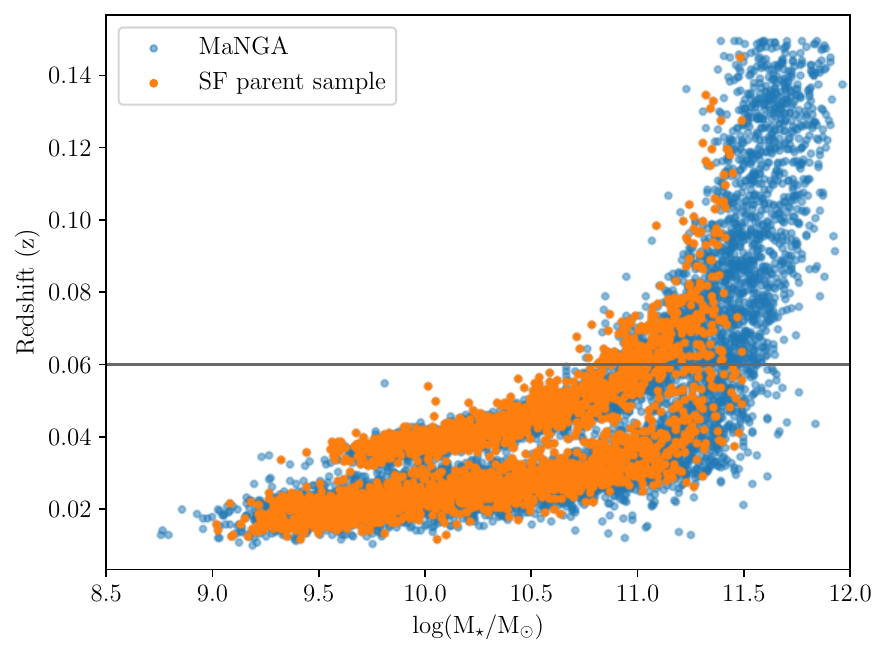}
    \includegraphics[width = 3.2 in]{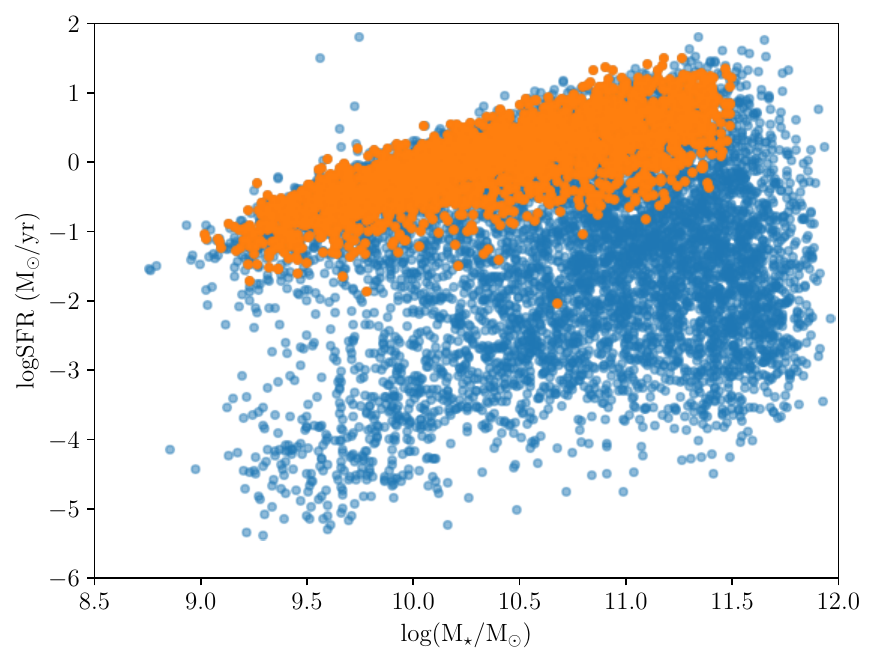}
    \caption{Our {\it SF parent sample} selection of $N = 2632$ galaxies (orange points) plotted alongside the whole of the MaNGA sample (blue points), showing redshift (upper) and SFR (lower; ({\tt logSFR\_Ha} from Pipe3D) as a function of stellar mass. We observe that our sample is from both the primary (lower redshift for mass) and secondary (higher redshift for mass) samples of MaNGA, and almost entirely in the star-forming sequence.}
    \label{fig: mass-SFR-z}
\end{figure}
\begin{enumerate} 
     \item {\it Possible to Measure Gas Phase Metallicity:} Galaxies in our sample must have emission line regions suitable for measuring the gas phase metallicity (see \S \ref{sec:methods}). We require the number of spaxels with SF ionized emission lines to be $N > 9$. This selection excludes galaxies that do not have SF regions. We use an analysis of the ionization source of emission lines through a Baldwin, Philips, and Terlevich \citep[BPT,][]{BPT1981} diagram in {\tt Marvin} \citep{Cherinka2019}. The {\tt Marvin} SF BPT classifications masks are based on the [NII]$\lambda6584$, [OI] $\lambda 6300$, and [SII] $\lambda \lambda 6717, 31$ criteria from \citet{Kewley2006} and the \citet{Kauffmann2003} classification. This is identical to the cut made by \citet{Belfiore2017} in their prior analysis on gas phase metallicity gradients with an earlier release of MaNGA data.\footnote{At an advanced stage of this manuscript we became aware of the work of \citet{Scudder2025} who check the reliability of metallicity measurements from MaNGA spectra adjacent to non-SF spaxels in the BPT diagram. We have not implemented their recommendations to remove a small fraction of SF spaxels adjacent to non-SF emission, but note that we primarilly use the 03N2 indicator, which they find is not significantly impacted.}
     \item {\it Mass Range:} $ 9.0 <\log (M_\star/M_\odot) < 11.5$: a similar selection to  \citet{Belfiore2017}\footnote{While \citet{Belfiore2017} also uses a range $ 9.0 <\log (M_\star/M_\odot) < 11.5$ their masses use a Chabrier IMF (the Pipe3D masses we use, which are based on a Salpeter IMF), and so are roughly 0.2dex (a factor of 0.63) larger; i.e. their limits correspond to $9.2 <\log (M_\star/M_\odot) < 11.7$ for Pipe3D masses.}, that eliminates mass bins that have too few galaxies for reliable population statistics. We use the mass estimated within the MaNGA bundle with the spectral fitting code {\tt Pipe3D} \citep{Sanchez2022}.
     \item {\it Isolated galaxies}: We wish to obtain a sample of isolated galaxies suitable for investigating process of internal (secular) radial migration. We make use the limits, $d_{\rm nn} > 0.2$ Mpc and $p_{\rm merger} < 0.4$ where $d_{\rm nn}$ is the projected distance to the first nearest neighbor from \citet{Argudo2015}\footnote{Specifically ``HDU2: LSS environment 1Mpc all" from the DR17 VAC}, and $p_{\rm merger}$ is the GZ merger vote fraction. These conservative cuts remove any galaxy which might be tidally impacted by near neighbors. The $p_{\rm merger}$ cuts was motivated in part by the morphological changes seen in close pairs by \citet{Casteels2013}. 
    \item {\it Possible to see internal structures:} We need to be able to see internal features for our study. So we make a selection on major-to-minor axis ratio $a/b>0.4$ to exclude highly inclined systems where the presence of spiral arms and/or galactic bars is harder to identify. We make use of {\tt ELPETRO$\_$BA} from NSA to implement this selection. 
\end{enumerate}

These four criteria create our {\it SF parent sample} of 2,632 galaxies that are isolated, face-on, have some SF emission line regions, and contain enough data to obtain metallicity trends (masking out low SNR and flagged spaxels). {\tt Marvin} allows users to implement a SNR cut based on the emission lines used to generate a BPT diagram for a given galaxy. We mask out spaxels with $\rm SNR < 3$ to match the {\tt Marvin} minimum threshold. The {\it SF parent sample} uses the same base criteria as \citet{Belfiore2017}, however we make use of the final MaNGA sample so our sample is three times as large. Additionally, we introduce the $d_{\rm nn} > 0.2$Mpc selection to select sufficiently isolated galaxies such that our results do not have any effects from tidal disruption or merging. Fig. \ref{fig: mass-SFR-z} shows the {\it SF parent sample} compared to the entire MaNGA sample with panels showing redshift and SFR as functions of stellar mass. As is clear in this figure, our {\it SF parent sample} is selected from both the Primary and Secondary MaNGA samples, and is predominantly in the star-forming region. 

In addition to expanding the sample size from 
\citet{Belfiore2017}, we are interested in probing various morphological characteristics as they play a role in the radial trends in gas phase metallicity. We introduce a redshift limit once we make cuts on morphology to exclude galaxies that may have features that are unresolved due to their redshifts. The redshift limit we maintain for all samples moving forward is $z < 0.06$ (gray horizontal line in top panel of Fig. \ref{fig: mass-SFR-z}), which corresponds to a physical resolution of $\sim$ 3 kpc. This limits the {\it SF parent sample} to $N = 2327$.
 \begin{figure*}
     \centering
     \includegraphics[width = 7.1 in]{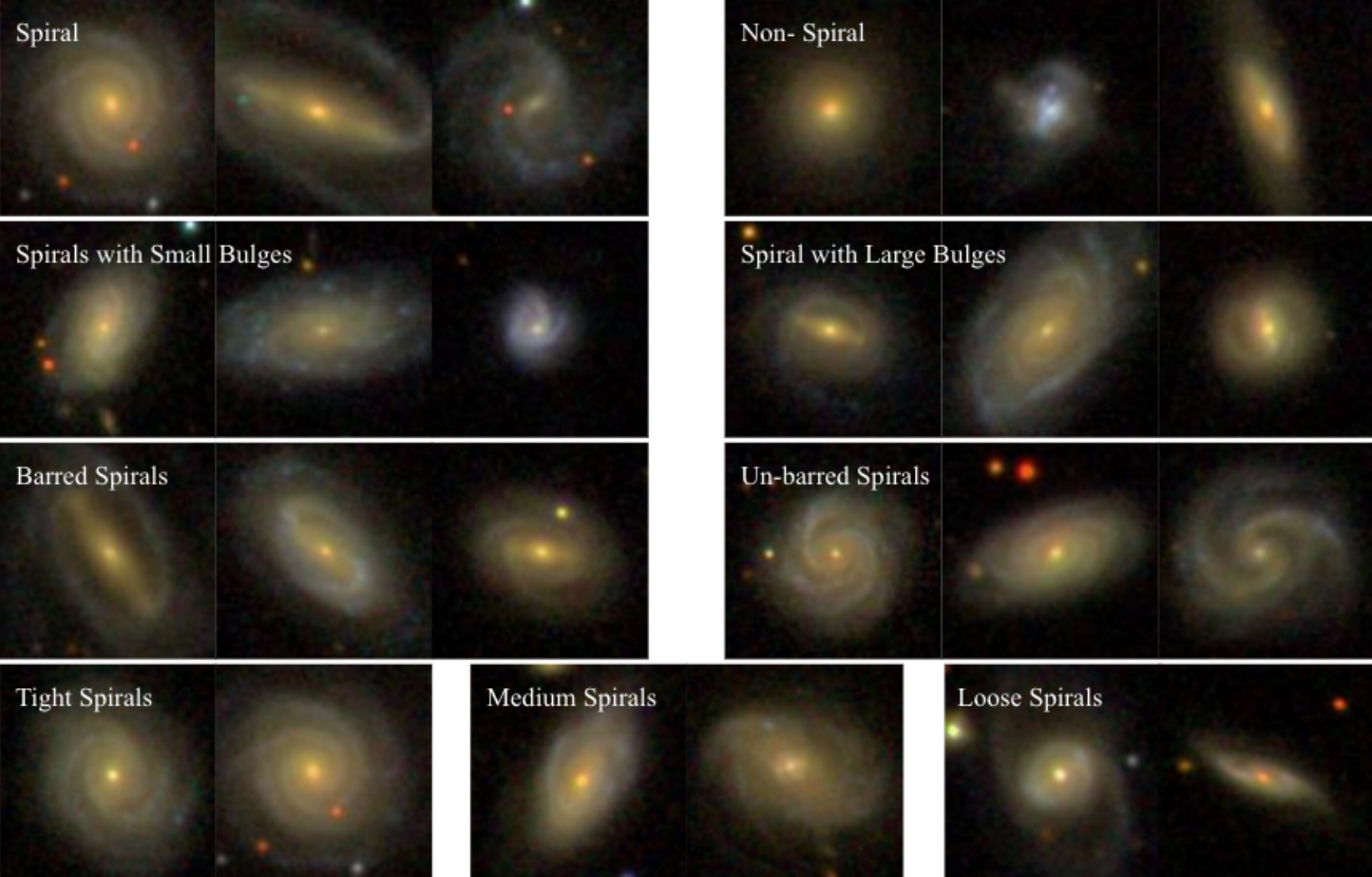}
     \caption{Example $gri$ images from morphological sub-samples.}
     \label{fig:example gals}
 \end{figure*} 
We are first interested in the impact of the presence of significant spiral structure. To separate our {\it SF parent sample} into a sample of galaxies that have distinct face-on spirals and galaxies without visible spirals, we use GZ morphology suggested thresholds for well sampled galaxies from \citet{Willett2013} as follows:
\begin{itemize}
    \item {\it spiral galaxy sample} - $p_{\rm features} \ge 0.5$, $p_{\rm notedgeon} > 0.715$ and $p_{\rm spiral} > 0.519$ ($N = 1233$) -- provides a sample of star-forming galaxies with clear spiral structure
    \item {\it non spiral galaxy sample} -- $p_{\rm features} \le 0.5$ OR $p_{\rm features} \ge 0.5$, with $p_{\rm notedgeon} > 0.715$ and $p_{\rm nospiral} > 0.519$ ($N = 878$) -- provides a sample of galaxies which are (in the GZ language) ``smooth" star-forming galaxies or ``featured" star-forming galaxies with no visible spirals. This is likely a mixture of blue ellipticals, S0s and irregular galaxies. This subsample could also include weak or unclear spirals including galaxies with flocculent spiral arms which may not always be visible in the SDSS images.
\end{itemize}
This selection removes 216 edge-on ($p_{\rm edgeon} > 0.29$ in GZ2) disc-like SF galaxies. 

Bulges in spiral galaxies are thought to have an impact on the formation of spiral galaxies and star formation processes and motions within the bulge. We investigate potential impacts of bulge prominence on the radial gas-phase metallicities of our spiral galaxy sample. There are many ways to characterize bulges sizes in spiral galaxies, but we continue to use GZ classification votes for consistency with other morphological indicators. From the spiral galaxy sample, we assign each galaxy a bulge prominence score using Equation 3 from \citet{Masters2019}:
\begin{equation}
    B_{avg} = 0.2p_{\rm just noticeable} + 0.8p_{\rm obvious} + 1.0p_{\rm dominant},
\end{equation}
which takes into account GZ volunteer answers to the question regarding the prominence of the central bulge size. Bulge sized subsamples are created using the following thresholds:
\begin{itemize}
        \item {\it small bulge sample} - $B_{avg} <= 0.4$ (N = 674) - spiral galaxies with visibly small bulges.
    \item {\it large bulge sample} - $B_{avg} > 0.4$ (N = 559) - spiral galaxies with visibly large bulges.
\end{itemize}
For a comparison of this measure of bulge size with more traditional $B/T$ measures, see \citet{Hart2017} and \citet{Masters2019}.

Bars in spiral galaxies have various impacts on the dynamical evolution of the arms and thus bars may also impact radial metallicity trends. To investigate this, we make subsets based on the presence or absence of a strong bar in galaxies in the {\it spiral galaxy sample}. These samples are created by using the following thresholds: 
\begin{itemize}
    \item {\it unbarred spiral sample} - $p_{\rm bar} < 0.3$ ($N = 664$) - a sample of spiral galaxies with no visible bar. 
    \item {\it barred spiral sample} - $p_{\rm bar} > 0.5$ ($N = 369$) - spiral galaxies with strong bars present. 
\end{itemize}
The remaining 200 galaxies in the spiral sample have intermediate values of $p_{\rm bar}$ which might indicate the presence of a weak bar, or uncertainty about the presence of a bar so are excluded from consideration.

For our final sub-sample selections, we investigate the impact of the level of arm winding. We focus on our unbarred spiral galaxy sample to separate the impact of a bar from the spiral arms alone. For each galaxy we calculate an arm winding score using the arm windiness score equation from \citet{Masters2019}:
\begin{equation}
    w_{\rm avg} = 0.5p_{\rm medium} + 1.0p_{\rm tight},
    \label{eq: winding}
\end{equation}
where $p_{\rm medium}$ is the vote fraction from GZ2 for the percentage of volunteers who selected medium for their answer to the arm winding level question, and $p_{tight}$ is the percentage of volunteers who selected tight as their answer. This score codes three possible answers into a single numerical range. A loose wound spiral with $p_{\rm loose}=1$ has $w_{avg} = 0$, while a tight wound spiral, with $p_{\rm tight}=1$ has $w_{avg} = 1$. These $w_{avg}$ values are converted into pitch angles using Equation 2 from \citet{Masters2019}, which was based on comparison with the sample discussed in \citet{Hart2017},
\begin{equation}
    \Psi = (25.6\pm0.5) - (10.8 \pm0.8)w_{avg},
    \label{eq:pitch angle}
\end{equation}
where $\Psi$ is the pitch angle in degrees. We bin the 664 unbarred spiral galaxies by arm winding levels, using $w_{avg} < 0.6$ for `loosely' wound spiral arms, and $w_{avg} > 0.8$ to define `tightly' wound spiral arms. These limits correspond to $\Psi > 19\pm 0.7 ^\circ$ (loose) and $\Psi < 17\pm 0.8 ^\circ$ (tight). Galaxies that fall between these thresholds are called `medium'. These thresholds create samples of 207 tightly wound, 286 medium wound, and 171 loosely wound unbarred spiral galaxies. Example images of galaxies from most of our subsamples are shown in Fig. \ref{fig:example gals}. 

\subsection{Computing Gas Phase Metallicity Trends}
\label{sec:methods}
In this work we primarily use the O3N2 metallicity calibration from \citet{PP2004}, which uses the $F[{\rm OIII} \lambda 5008]$, $F[{\rm H\beta} \lambda 4862]$, $F[{\rm NII} \lambda 6585]$, and $F[{\rm H\alpha} \lambda 6564]$ fluxes to calculate the O/H metallicity as 
\begin{equation}
\begin{split}
    &12 + \log{\rm (O/H)} = \\
    &8.73 - 0.32 \log\left(\frac{{F[{\rm OIII} \lambda 5008]}/{F[{\rm H\beta} \lambda 4862]}}{{F[{\rm NII} \lambda 6585]}/{F[{\rm H\alpha} \lambda 6564]}}\right).
    \label{eq:O3N2metallicity}
\end{split}
\end{equation}
We also make use of the N2 index metallicity calibrator from \citet{PP2004} which uses the ratio of the [NII] and H$\alpha$ fluxes as an estimate of metallicity via 
\begin{equation}
    12 + \log{\rm (O/H)} = 8.90 + 0.57 \log\left(\frac{F[{\rm NII} \lambda 6585]}{F[{\rm H\alpha} \lambda 6564]}\right).
    \label{eq:N2metallicity}
\end{equation}
Finally, we check results using a third metallicity calibrator based on the $R_{23}$ index given in \citet{Pagel1979}, or 
\begin{equation}
    R_{23} = \frac{(F[{\rm OII}\lambda 3727] + F[\rm{OIII}\lambda 4959] + F[{\rm OIII} \lambda 5007])}{F[{\rm H\beta} \lambda 4862]}.
    \label{eq:r23}
\end{equation}
The $R_{23}$ relation is double valued, however based on estimates from the other metallicity measures, the galaxies in our sample reside exclusively on the upper branch. Thus, we use the analytical fit to the upper branch of the $R_{23}$ metallicity relation described in \citet{Tremonti2004} to recover the $\rm 12 + log(O/H)$ metallicity value via 
\begin{equation}
    12 + \log{\rm (O/H)} = 9.185 - 0.313x - 0.264x^2 - 0.321x^3,
    \label{eq:r23_fit}
\end{equation}
where $x\equiv \log R_{23}$. 

We dust correct all emission line fluxes using the Balmer decrement. We calculate the Balmer decrement for each galaxy as given in \citet{momcheva2013} \footnote{We use the Balmer decrement function as written by X. Prochaska in the $\rm F^4$ github, found at: https://github.com/FRBs/}. The decrement is then used to correct all emission lines using the correction equation:
\begin{equation}
    {\rm F_{int}} = {\rm F_{obs}}^{0.4k{\rm E(B-V)}},
\end{equation}
where {\it k} is the extinction curve given in \citet{fitzpatrick2007}. 

A review of current metallicity calibrators \citep{Kewley2019} shows that strong line calibrators can have certain limitations. Fig. 8 from this review shows that the sample of calibrator methods for O3N2 and N2 will result in a range of metallicity values with a scatter of up to one dex. We select the calibrators for each method that best fit the median of the possible range of values (given by Eqs. \ref{eq:O3N2metallicity} and \ref{eq:N2metallicity}). In what follows we will check how results might change using different calibrators.

\begin{figure}
    \centering
    \includegraphics[width = 3.2 in]{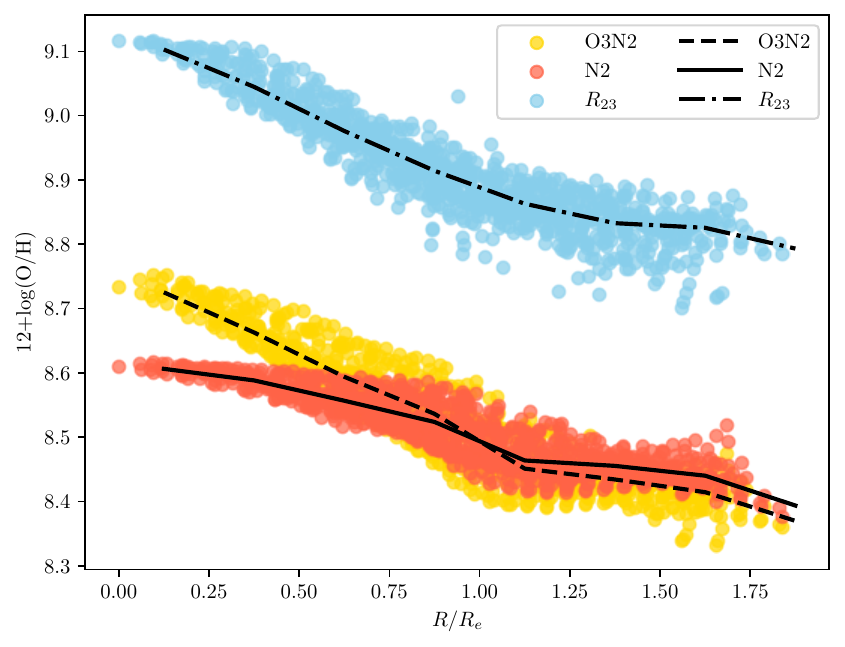}
    \caption{Radial trend of metallicity using all three emission line metallicity indicators for example galaxy, MaNGA-ID: 1-62035 (see Figure \ref{fig:examplegalaxyanddata}); a loosely wound spiral galaxy in our sample. Points show individual spaxel measurements of gas-phase metallicity using the O3N2 (yellow), the N2 (red), and the R23 indicators (blue). Over-plotted on each is line showing the binned average for this galaxy.}
    \label{fig:exampletrends}
\end{figure}

For each galaxy in our \emph {SF parent sample}, we use {\tt Marvin} \citep{Cherinka2019} to obtain the relevant emission line flux map values necessary to apply Equations \ref{eq:O3N2metallicity}, \ref{eq:N2metallicity}, and \ref{eq:r23_fit} and estimate gas-phase metallicities. A metallicity value is calculated for emission line measures in each spaxel in each galaxy and for each metallicity calibrator. The values for each calibrator are then binned and averaged radially (using the deprojected radius values based on the thin-disc approximation available in MaNGA map files\footnote{i.e. {\tt spx\_ellcoo\_r\_re}}) such that we obtain a radial metallicity trend for each galaxy in our \emph {SF parent sample}. We create up to 35 radial bins equally spaced across 0--3 $R_e$ (not all galaxies extend beyond $1.5R_e$).  An example of the three radial trends (for the three gas-phase metallicity indicators we discuss) is shown in Figure \ref{fig:exampletrends} (this figure shows radial trends for the same galaxy as we show example MaNGA maps for in Figure \ref{fig:examplegalaxyanddata}). 

\begin{figure*}
    \centering
    \includegraphics[width = 7.1 in]{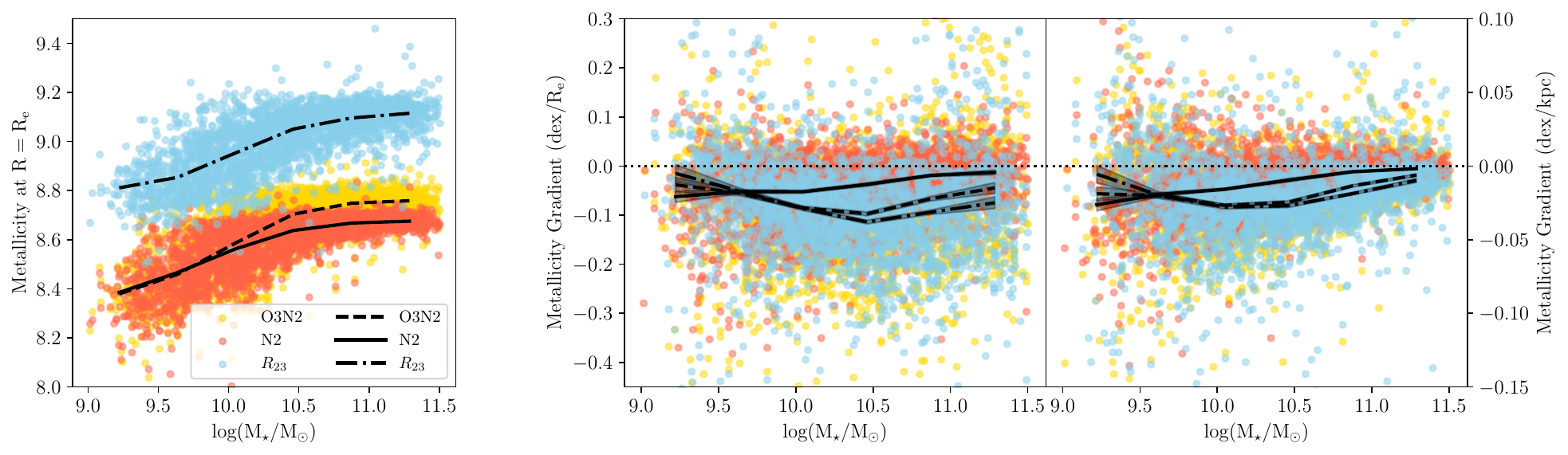}
    \caption{Gradients (middle; in dex/$R_e$; right: in dex/kpc) and intercepts (left; Z = $12+\log$(O/H)) of straight line fits to radial trends of all galaxies in the \emph {SF parent sample} as a function of stellar mass for the O3N2 index (yellow), N2 index (red), and $R_{23}$ index (blue) based metallicity calibrators used in this paper. Binned averages are shown in black for all calibrators (dashed line for O3N2, solid for N2, and dot-dashed for $R_{23}$) with the grey shaded region indicating $\sigma/\sqrt{N}$ errors on the binned average.}
    \label{fig:grad slope mass all}
\end{figure*}

 In addition to generating a binned average trend per galaxy, we also perform a simple linear regression to each galaxy's spaxel based metallicity data across 0.5--3 $R_e$. We exclude the first 0.5 $R_e$ from the linear trend fit to avoid possible effects of the bulge and central spaxels. We show the results from these fits to all galaxies in the \emph{SF parent sample} and all three metallicity indicators in Figure \ref{fig:grad slope mass all}. We show metallicity at $R/R_e$ (left), gradient of the linear fit in units of both dex/$R_e$ (middle) and dex/kpc (right) as functions of stellar mass.  While each gas-phase calibrator shows slightly different results we find that with increasing galaxy mass, the metallicity values at $R = R_e$ increase, as expected from the MZR. We also notice that the gradient values are quite  similar in all metallicity indicators (albeit with a large scatter).
 
The $R_{23}$ calibrator clearly results in systematically higher metallicity values at all radii for all of our samples, however we otherwise obtain comparable population wide results. Comparing the line fits to the metallicity trends from the different indicators, we find that a Spearman's rank correlation test gives a correlation coefficient of 0.668 for relationship between the gradients measured in the O3N2 and N2 indicators and 0.728 for the metallicity values at $R = R_e$. They are considered well correlated ($p$-value = 0.00). The O3N2 calibrator compared to the $R_{23}$ calibrator has correlation coefficients of 0.750 and 0.884 respectively, also with p-values of 0.0.

Thus, within our \emph {SF parent sample}, we conclude that we obtain comparable population wide results using all three metallicity calibrators. The different indicators return different values for metallicity with some significance, but the broad trends seen in the shapes of radial profiles match reasonably well.  The source for any minor differences between the outcome from these metallicity calibrators is discussed extensively in \citet{Kewley2019}. Conversions between these, and various other metallicity indicators possible with MaNGA data are provided in \citet{Scudder2021}. Moving forward, we present population wide trends calculated using the O3N2 index based values, although we check that there are not qualitative differences in our results which depend on the choice of indicator.

\subsection{Sub-Sample Averages}
 In each sub-sample of interest (as defined in Section \ref{sec:sample selection}), a binned statistic trend is computed averaging the gas-phase metallicity as a function of radius ($R_e$) for all the galaxies in the sub-sample. We do this only for radial bins with data from $N > 12$ galaxies so that the median calculated for each radial bin is not impacted by small number statistics. This choice of minimum bin size matches that used by \citet{Belfiore2017}.

 \subsection{Controlling for Mass Dependence} \label{sec:massmatch}
 We are interested primarily in the impact internal structures have on metallicity gradients. These structures are known to have dependence on the total mass of the galaxy (e.g. \citealt{Nair2010} show how the galactic bar fraction depends on stellar mass), which therefore may act as a confounding variable due to the MZR, \citep{Tremonti2004}. For the full \emph {SF parent sample} it is possible create very narrow mass bins of 0.25 dex and plot radial gas-phase metallicity trends in each of these separately, to account for this expected mass dependence. For all subsequent sub-samples, the smaller sample sizes makes this technique impossible. 
 
\begin{figure*}
    \centering
    \includegraphics[scale = 0.5]{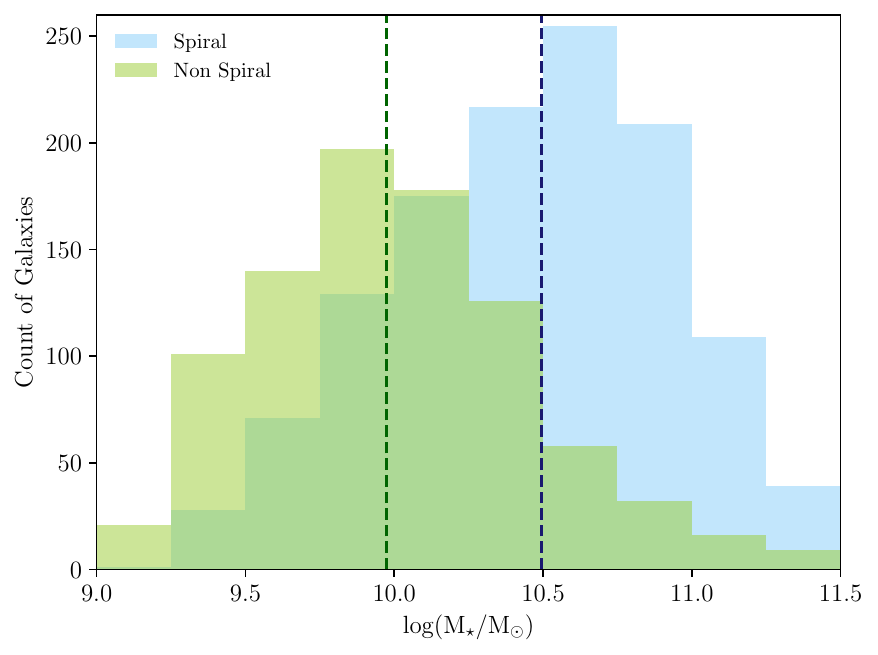}
    \includegraphics[scale = 0.5]{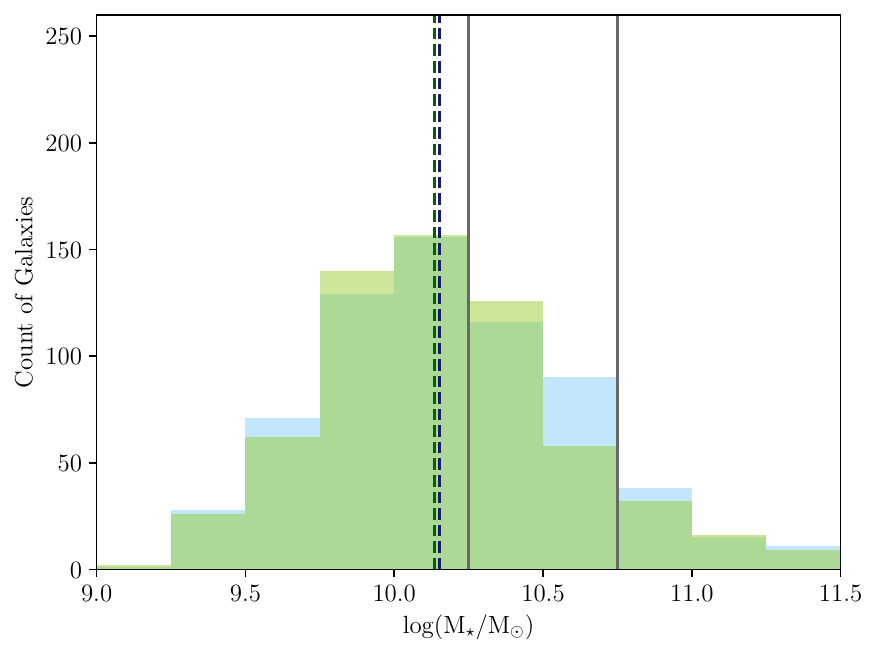}\\
    \includegraphics[scale = 0.5]{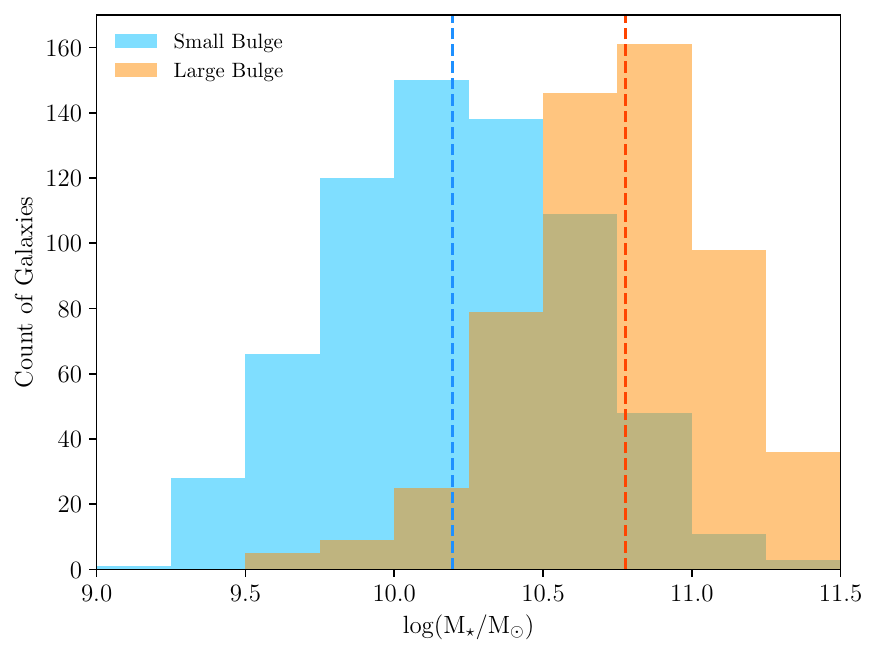}
    \includegraphics[scale = 0.5]{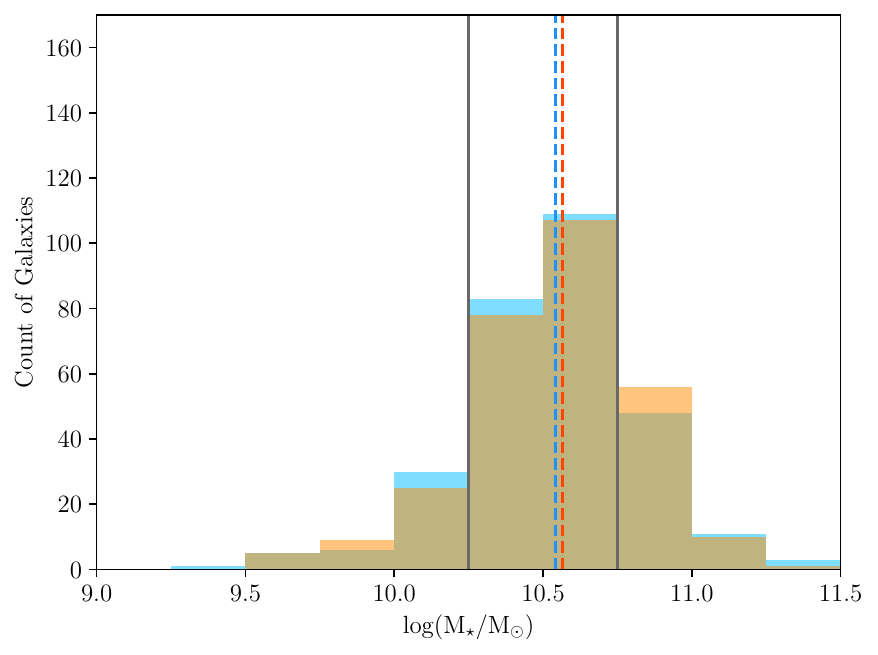}\\
    \includegraphics[scale = 0.5]{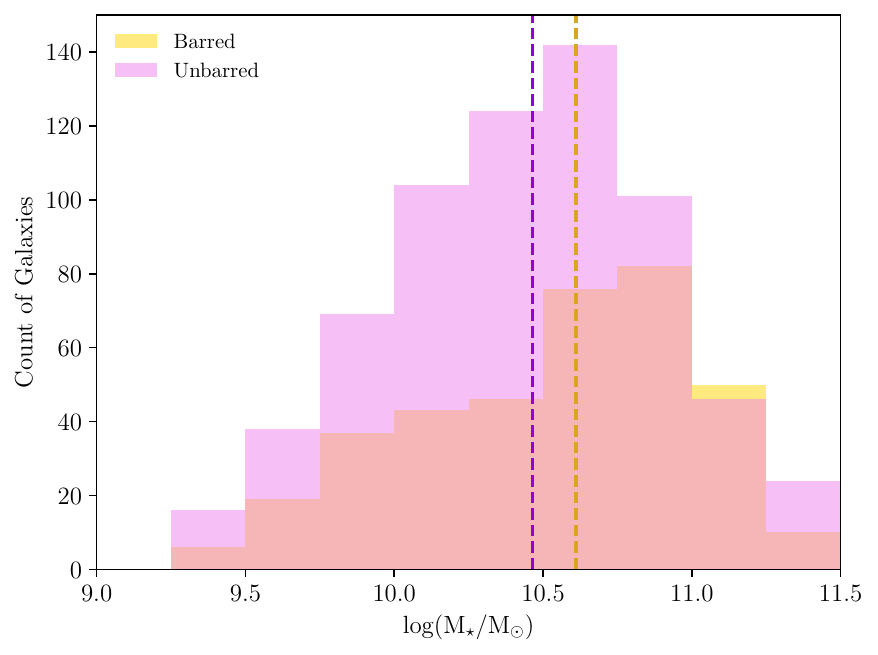}
    \includegraphics[scale = 0.5]{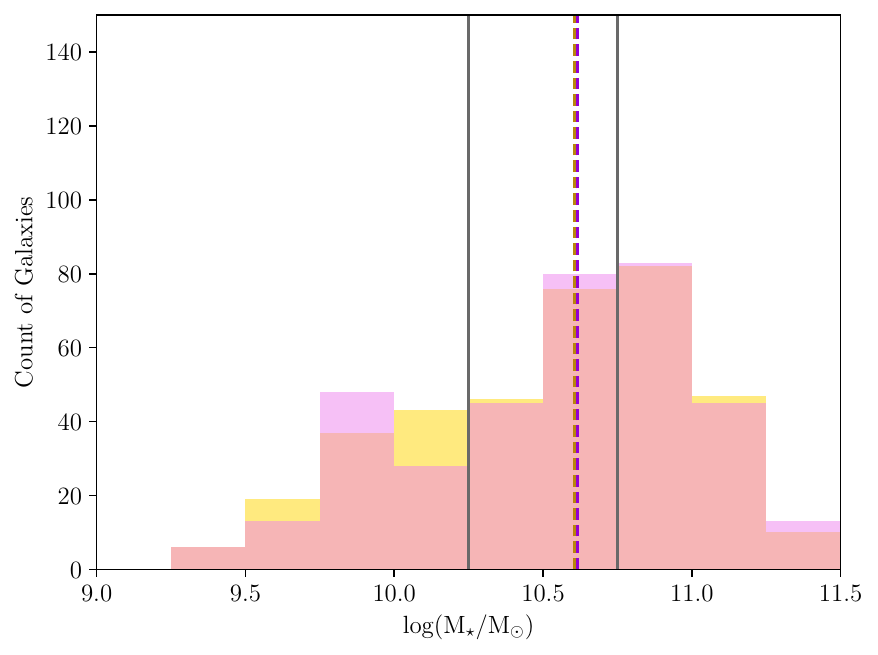}\\
    \includegraphics[scale = 0.5]{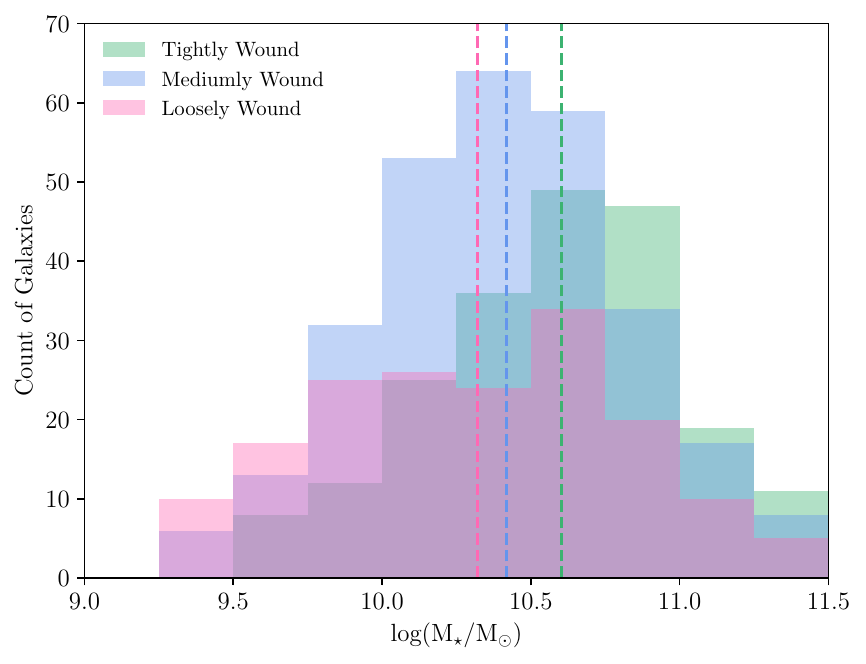}
    \includegraphics[scale = 0.5]{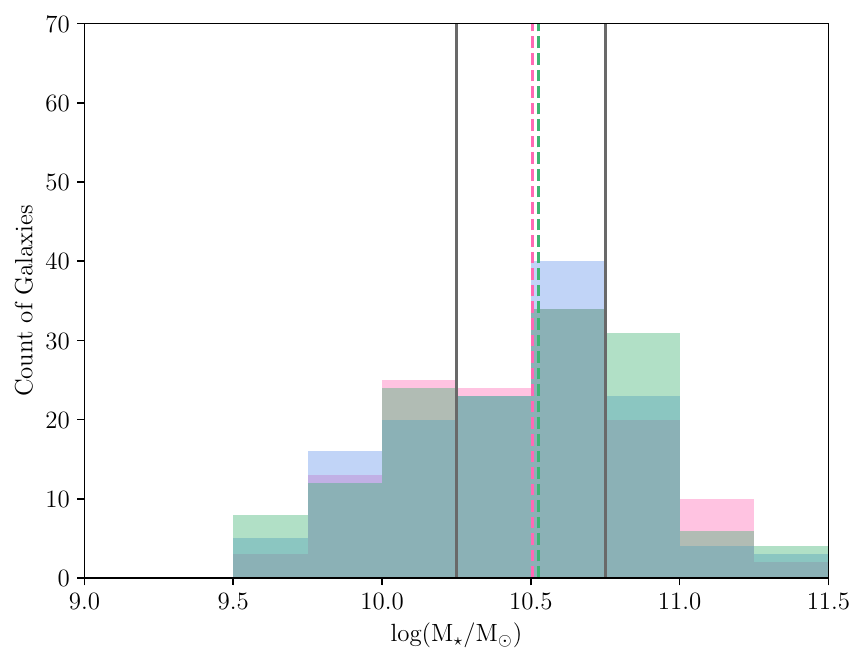}
    \caption{Normalized mass distributions of the samples before (left) and after (right) mass matching. Spiral (sky blue) and non-spiral (green) samples shown in the top row. Small bulge (blue) and large bulge (orange) samples are shown in second row. Barred (yellow) and unbarred (violet) spiral samples shown in third row. Tight (light green), medium (light blue) and loosely (pink) wound unbarred spiral galaxy samples shown in bottom row. The dashed vertical lines show the median mass value for each of the samples and the vertical gray lines indicate the thresholds for mass binning for all samples. Sample sizes before and after mass matching are given in Table \ref{tab:samples}.}
    \label{fig:mass}
\end{figure*}

 For the smaller sub-samples, we use broad mass bins, with thresholds at $\log (M_\star/M_\odot) = 10.25$ and $\log (M_\star/M_\odot) = 10.75$, and mass-match each sub-sample. The mass matching algorithm assigns a weight to each galaxy in the smallest of the samples and selects down from both to create equivalently weighted mass distributions. Histograms of the mass distribution of our subsamples before and after mass matching are shown in Figure \ref{fig:mass}, and the sizes of the resulting samples are given in Table \ref{tab:samples}.

\begin{table}
\centering 
\begin{tabular}{lrrrr}
\hline
\textbf{Sample}&\textbf{All~~~}&&$\log (M/M_\odot)$\\
\textbf{}&\textbf{}&\textbf{$<$10.25}&\textbf{Mid~~~~}&\textbf{$>$10.75}\\
\hline 
\hline
All SF & 2,632~~~~\\
\hline
Non-spiral & 878(628) & 637(387) & 184(184) & 57(57)\\
Spiral & 1233(655) & 404(385) & 472(206) & 357(64)\\
\hline
~Small Bulge & 674(296) & 365(42) & 247(192) & 62(62) \\
~Large Bulge & 559(291) & 39(39) & 225(185) & 295(67) \\
\hline
~Barred & 369(366) & 105(105) & 122(122) & 142(139) \\
~Unbarred & 664(361) & 227~(95) & 266(125) & 171(141) \\
\hline
~~~Tight & 207(142) & 45~(44) & 85~(57) & 77~(41) \\
~~~Medium & 286(134) & 104~(41) & 123~(63) & 59~(30) \\
~~~Loose & 171(131) & 78~(41) & 58~(58) & 35~(32) \\
\hline
\end{tabular}
\caption{Sub-sample sizes. We provide sample sizes both before and after (in parentheses) the mass-matching process. Mass matching randomly down selects over-represented mass ranges from the larger of the two samples being compared. Figure \ref{fig:mass} shows histograms of the distributions of stellar masses before and after matching. \label{tab:samples}}
\end{table}

 The broad mass bins were selected to best characterize the spiral sub-samples to investigate how spiral arm morphology and the presence of a bar impact gradients (see lower two rows of Figure \ref{fig:mass}). Our sample of non-spiral star forming galaxies (\emph{no-spiral sample}) is found to skew to significantly lower mass than the star-forming spiral sample, which contains more high mass galaxies.
 
 The original and mass-matched mass distributions for all samples are shown in Fig. \ref{fig:mass}, with corresponding color coordinated vertical dashed lines displaying the median mass of each sample. For each of the three mass bins for the mass-matched samples that are being compared, binned statistic radial metallicity trends are computed and a simple linear regression is fit to the averaged trends. 
 
\section{Results} 
\label{sec:results}
In this section we will present all results obtained from the various samples to investigate the role of internal disc morphology - the \emph {SF parent sample}, \emph{spiral--non-spiral sample}, \emph{bulge size samples}, \emph{barred--unbarred sample}, and finally unbarred spirals by arm winding. We show results only for the O3N2 calibrator however we do check that we obtain similar plots for both N2 and R23 calibrators (see Section \ref{sec:methods} for more discussion of this).

\subsection{Gas Phase Metallicity Gradients in all SF Galaxies}\label{sec:parent}
\begin{figure}
    \centering
    \includegraphics[width = 3.7 in]{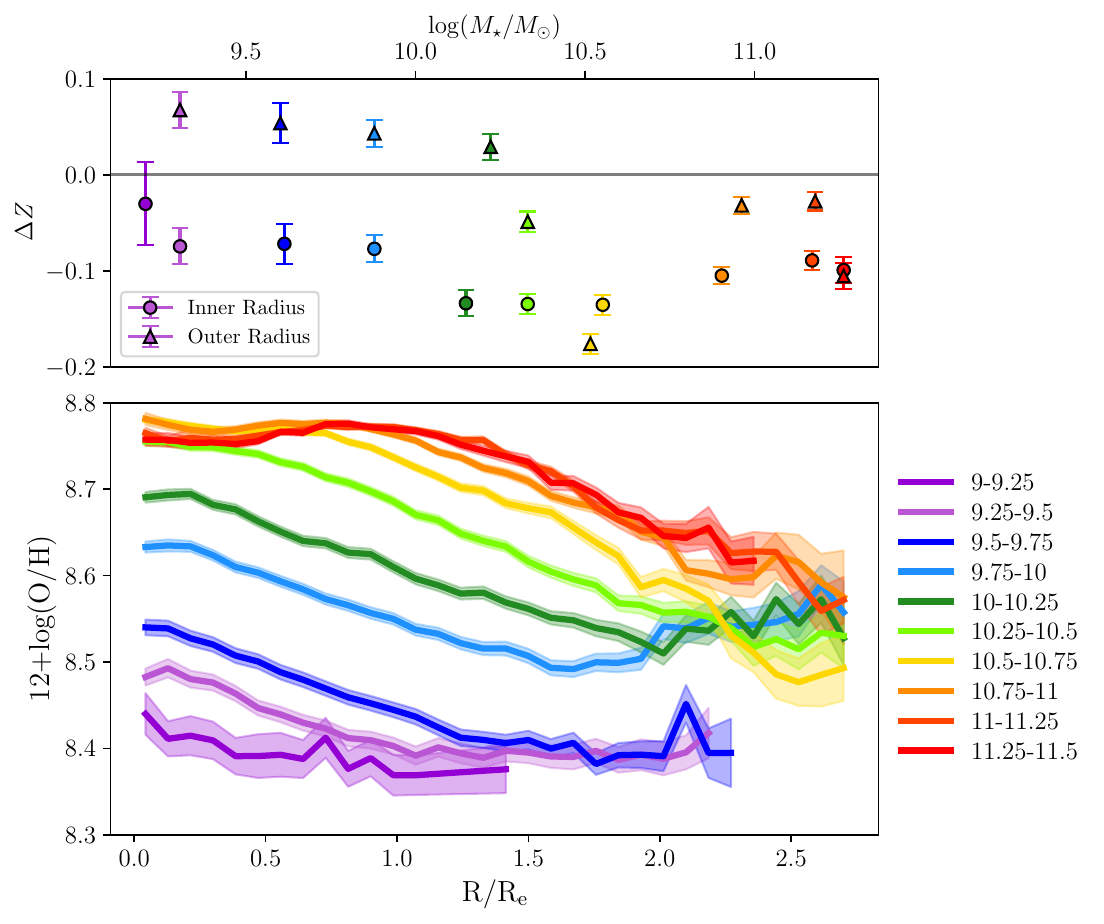}
    \caption{ \emph{Lower panel:} Binned gas-phase O3N2 index metallicity trends for 2632 galaxies in stellar mass subsamples (see legend for the $\log(M/M_\odot)\pm0.25$ dex mass ranges).  \emph{Upper panel:} Median change in metallicity ($Z = 12 + \log (O/H)$) in each subset between $0.6 R_{e}$ and $1.8 R_{e})$ (circles) or $1.8 R_{e}$ and $2.4 R_{e}$ (triangles). Points are plotted at the median mass in each mass range using the same color code as for the lower panel. The error bars in the upper panel and filled regions in the lower panel indicate the standard error on the mean, or $\sigma/\sqrt{n}$, where $\sigma$ is standard deviation, and $n$ is the number of galaxies for which a metallicity was calculated in that radial bin.}
    \label{fig:SF parent sample trends}
\end{figure}

We start by looking at how radial trends depend on stellar mass in our \emph {SF parent sample} of 2,632 isolated SF galaxies selected from the full MaNGA sample (see Figure \ref{fig:SF parent sample trends}). As was observed in \citet{Belfiore2017} for a earlier subset of MaNGA (see their Figure 3) we observe both increased average gas-phase metallicity and an increasing steepness (i.e. a more negative gradient, or a larger decrease from center to outskirts)  of the metallicity trends with galaxy mass, particularly in the outer parts of the galaxies. The larger error bars at $R\ge 2.5R_e$ reflect the increased uncertainty at larger radii due to fewer galaxies in the MaNGA sample with data in those radial bins. 

We note that these trends are not well fit by linear relations. We characterize the flattening of the gradients for each of these narrow mass bins in the upper panel of Fig. \ref{fig:SF parent sample trends} by showing the change in metallicity between a fixed `inner radii' and `outer radii.' We define `inner change' to be the change in metallicity between 0.6 $R_{e}$and 1.8 $R_{e}$ and `outer change' as the change between 1.8 $R_{e}$ and 2.4 $R_{e}$. Note that the lowest mass bin from 9--9.25 $\log (M_\star/M_\odot)$ does not have coverage out to 2.4 $R_{e}$, so we are unable to quantify the change for the outer parts for that mass bin. While there is an overall decline in metallicity with radius between 0.6-2.4 $R_{e}$ in all mass ranges, we measure slightly positive changes for the outer parts of the lowest mass galaxies. There is a clear transition around $\log (M_\star/M_\odot) = 10.3$ to negative metallicity changes for both inner and outer regions. Averaged over the whole radial range this results in a steepening (making more negative; or a larger negative change) of the overall trend with galaxy mass in the mass range below $\log (M_\star/M_\odot) = 10.3$, while for more massive galaxies, the overall gradients become flatter again.

\subsection{The Effect of Spiral Arms}
\begin{figure*}
    \centering
    \includegraphics[width = 7.1 in]{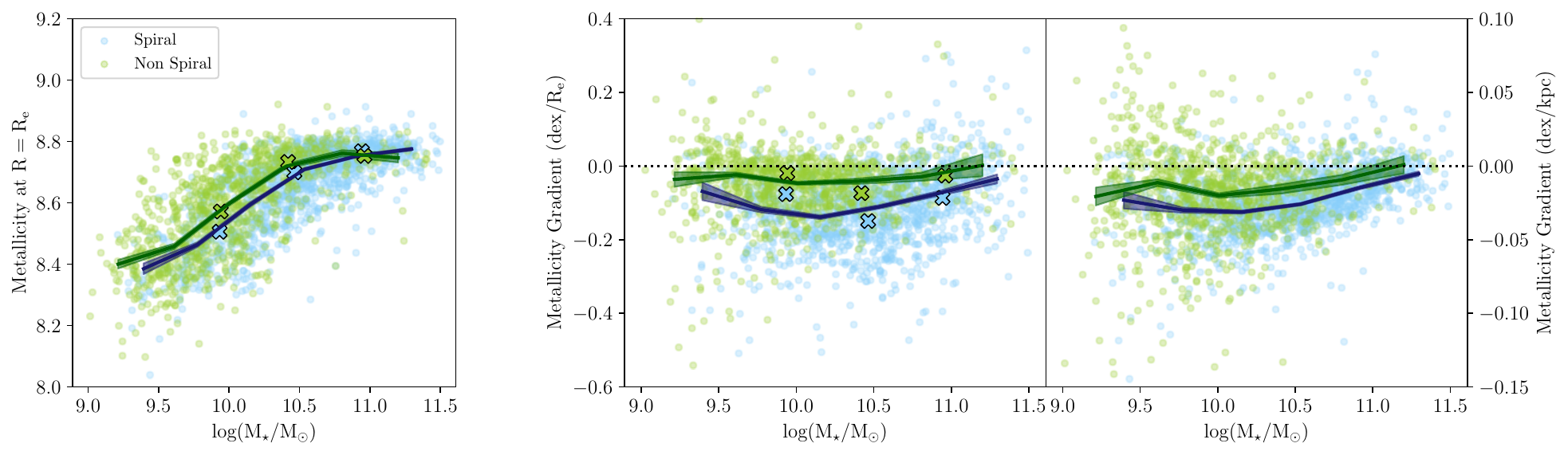}
    \caption{Gas phase metallicity values from O3N2 at $R = R_e$ (left) and gradients (middle; in units of dex/$R_e$; right: in units of dex/kpc) for straight line fits to the radial metallicity trends of all galaxies in \emph{spiral galaxy sample} (blue) and \emph{non-spiral galaxy sample} (green) as a function of stellar mass. Binned statistic trends to these relationships are shown for both samples, and the errors on these trend are shown in the shaded region. The values from linear regression fits to trends in broad mass bins (see Figure \ref{fig:disk v spiral trends mass matched})  are over-plotted as crosses in the corresponding colors. These data show that galaxies with visible spiral arms have slightly lower average metallicity, and flatter metallicity gradients.}
    \label{fig:spiral no spiral grad offset}
\end{figure*}

The \emph{SF parent sample}, is separated into a \emph{spiral sample} of 1233 spiral galaxies and a \emph{non-spiral sample} of 878 galaxies. In Figure \ref{fig:spiral no spiral grad offset} we show metallicity values at $R = R_e$ (left) and metallicity gradients (right) from the straight line fits to each galaxy's individual radial metallicity trend in these two sub-samples. We see only a small average difference in the metallicity values at $R = R_e$. These values increase with galaxy mass for both the \emph{spiral sample} and the \emph{non-spiral sample}, and are found to be very slightly higher for the  \emph{non-spiral sample}. However the gradient values for the \emph{non-spiral sample} are found to be flatter than those for the \emph{spiral sample} at all masses (the average gradient for all galaxies in the \emph{non-spiral sample} is $-0.016\pm0.001$~dex/kpc, compared to $-0.023\pm0.001$~dex/kpc for the \emph{spiral sample}). This is a clear signature that \emph{the presence of spiral arms correlates with steeper gas-phase metallicity gradients} in star forming galaxies. 

\begin{figure}
    \centering
    \includegraphics[width = 3.2 in]{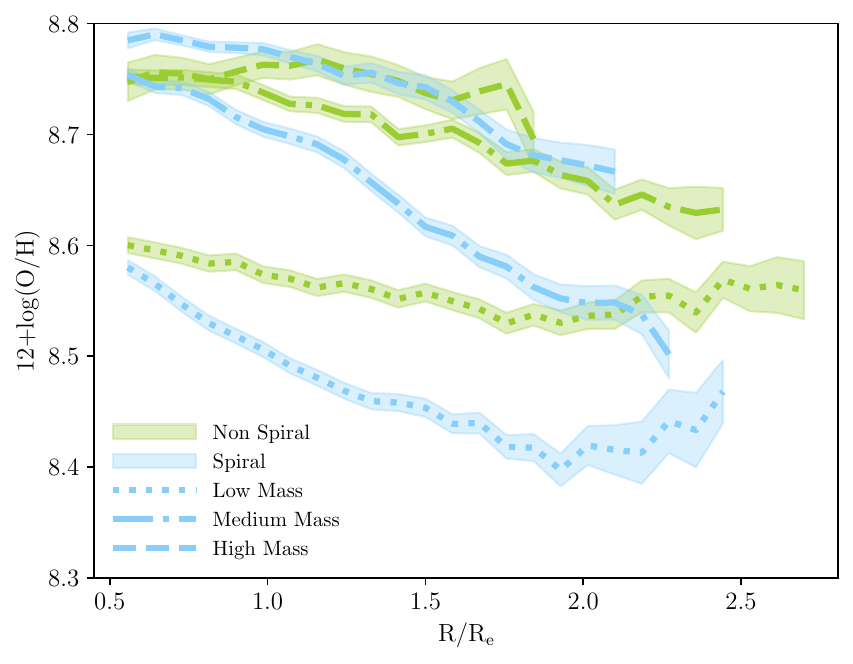}
    \caption{The O3N2 gas-phase metallicity trends for non-spiral galaxies (green) and visible spiral galaxies (blue) in sub-samples of three mass bins (with thresholds at $\log (M_\star/M_\odot) = 10.25$ and $\log (M_\star/M_\odot) = 10.75$; the line-styles indicate the different mass bins as shown in the legend).}
    \label{fig:disk v spiral trends mass matched}
\end{figure}

Figure \ref{fig:disk v spiral trends mass matched} shows these radial trends for the (mass-matched) spiral and non-spiral galaxy sub-samples in our three fixed mass bins (with thresholds at $\log (M_\star/M_\odot) = 10.25$ and $\log (M_\star/M_\odot) = 10.75$). In all three fixed mass bins, but especially in the two lower mass $\log (M_\star/M_\odot) < 10.75$ sub-samples, the \emph{non-spiral sample} shows flatter gradients than the \emph{spiral sample} radial trends, and that \emph{this trend of steeper gradients in SF galaxies with visible spirals is primarily driven by lower metallicities in the outer parts of the galaxies.}

  \subsection{The Effects of Bulge Size}
\begin{figure*}
    \centering
    \includegraphics[width = 7.1 in]{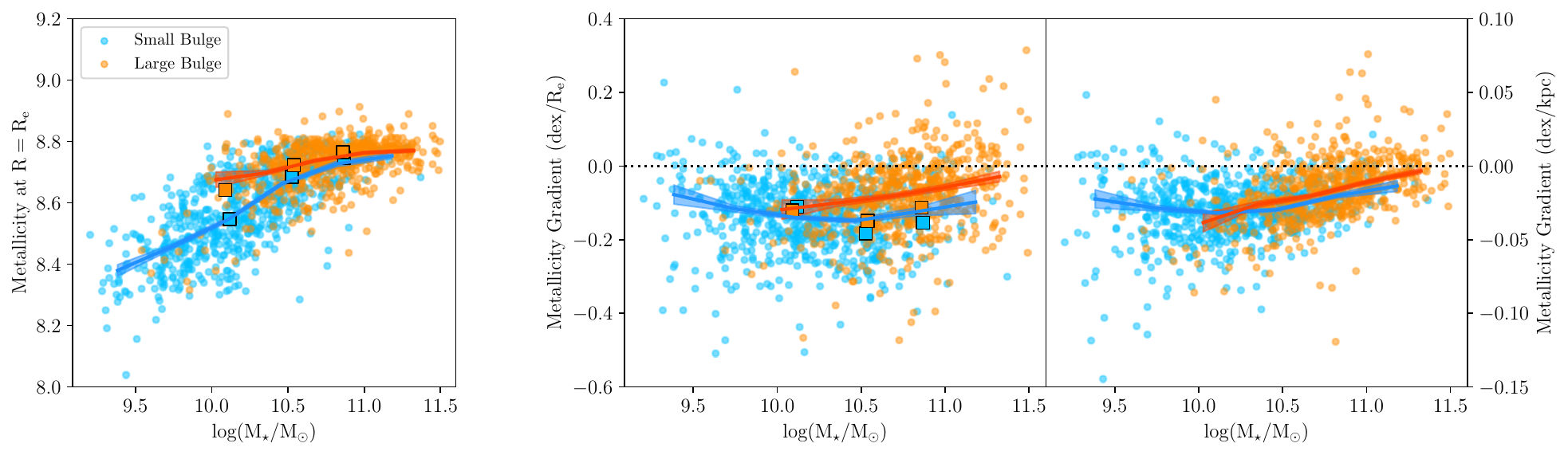}
    \caption{Metallicity values at $R = R_e$ (left) and gradients (middle and right) of straight line fits to all galaxies with large or small bulges from the spiral galaxy sample as a function of stellar mass. Binned statistic trends fit to these relationships are shown for both samples (blue for small bulges, and orange for large bulges), and the errors are shown in the shaded regions. Values from linear regression fits to trends Fig. \ref{fig:bulge size trends mass matched} are over-plotted in squares of corresponding colors to show agreement with mass-separated population trends.}
    \label{fig:bulge size slope int}
\end{figure*}

We separate our {\it spiral sample} into two broad categories based on bulge prominence, as defined in Section~\ref{sec:sample selection}, where each spiral galaxy is assigned to either the `small bulge' or `large bulge' samples. The metallicity values at $R = R_{e}$ (left) and gradients (right) from linear trends fit to each galaxy in the sample are shown in Figure \ref{fig:bulge size slope int}. {\it Spiral galaxies with larger bulges have, on average, higher metallicity values at all masses and shallower gradients than those with smaller bulges.} Specifically spirals with large bulges have an average gradient of $-0.016\pm0.001$~dex/kpc, compared to $-0.029\pm0.001$~dex/kpc for the small bulge spirals, however this may be mostly driven by the different mass ranges.

\begin{figure}
    \centering
    \includegraphics[width = 3.2 in]{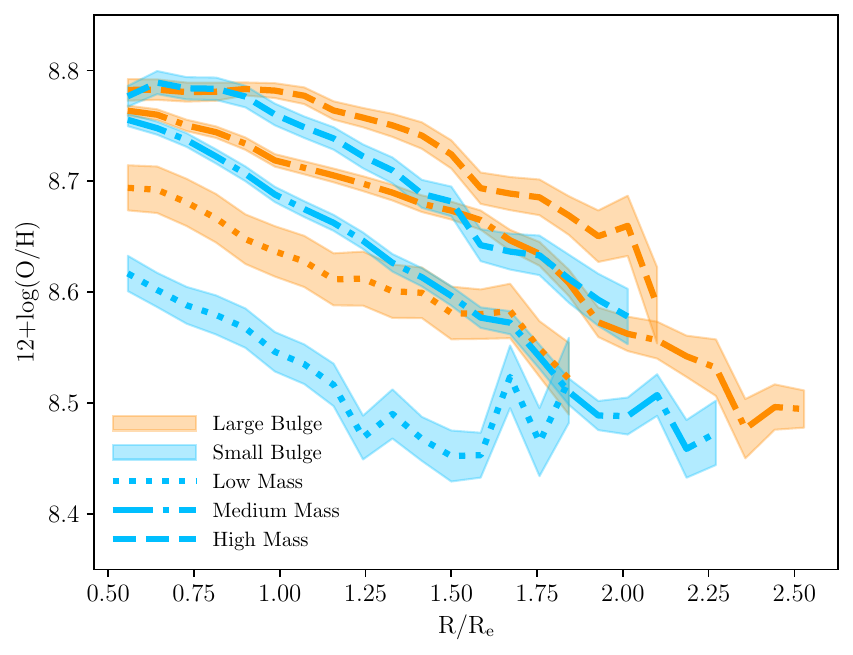}
    \caption{The O3N2 gas-phase metallicity trends for small bulges (blue) and large bulges (orange) in sub-samples of three mass bins (with thresholds at $\log (M_\star/M_\odot) = 10.25$ and $\log (M_\star/M_\odot) = 10.75$; the line-styles indicate the different mass bins as shown in the legend).}
    \label{fig:bulge size trends mass matched}
\end{figure}

After mass matching the small and large bulge samples within the three mass thresholds, we can still observe a slight effect (see Figure \ref{fig:bulge size trends mass matched}). We see a larger offset in the metallicity values in the lowest mass bin. For all three mass bins, the spiral galaxies with large bulges have shallower gradients and higher overall metallicity values. 

\subsection{The Effect of Bars}
\begin{figure*}
    \centering
    \includegraphics[width = 7.1 in]{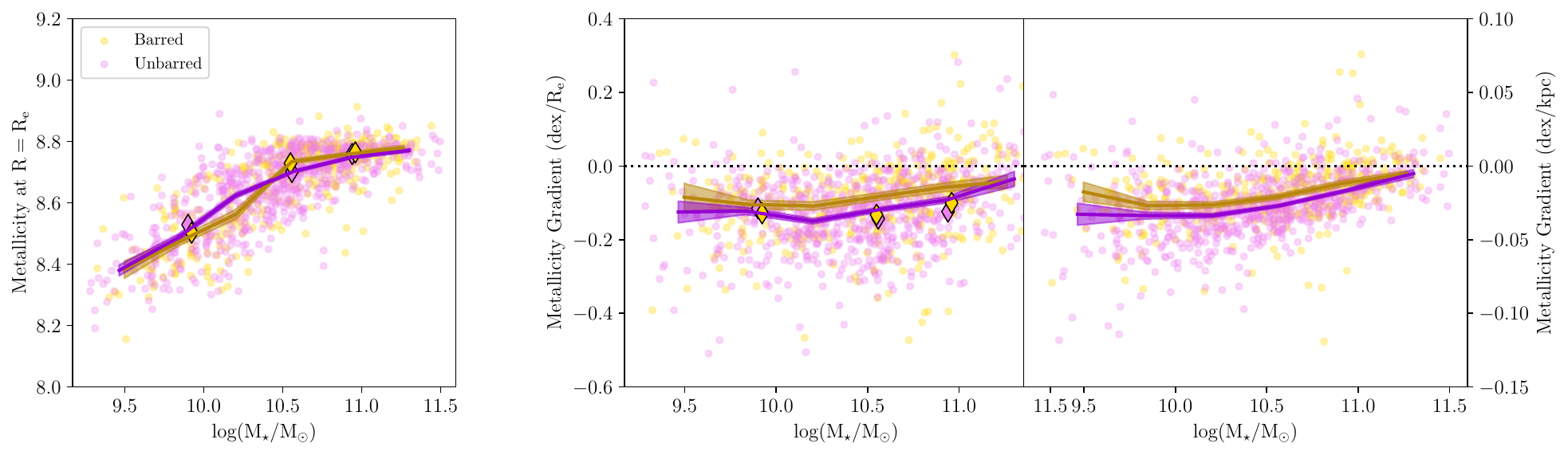}
    \caption{Metallicity values at $R = R_e$ (left) and gradients (middle and right) of straight line fits to all galaxies with strong bars or no bar from the spiral galaxy sample as a function of stellar mass. Binned statistic trends fit to these relationships are shown for both samples (yellow for bars, and pink for no bars), and the errors are shown in the shaded regions. Values from linear regression fits to trends Fig. \ref{fig:o3n2 barred unbarred trends} are over-plotted in squares of corresponding colors to show agreement with mass-separated population trends.}
    \label{fig:bar slope int}
\end{figure*}

\begin{figure}
    \centering
    \includegraphics[width = 3.2 in]{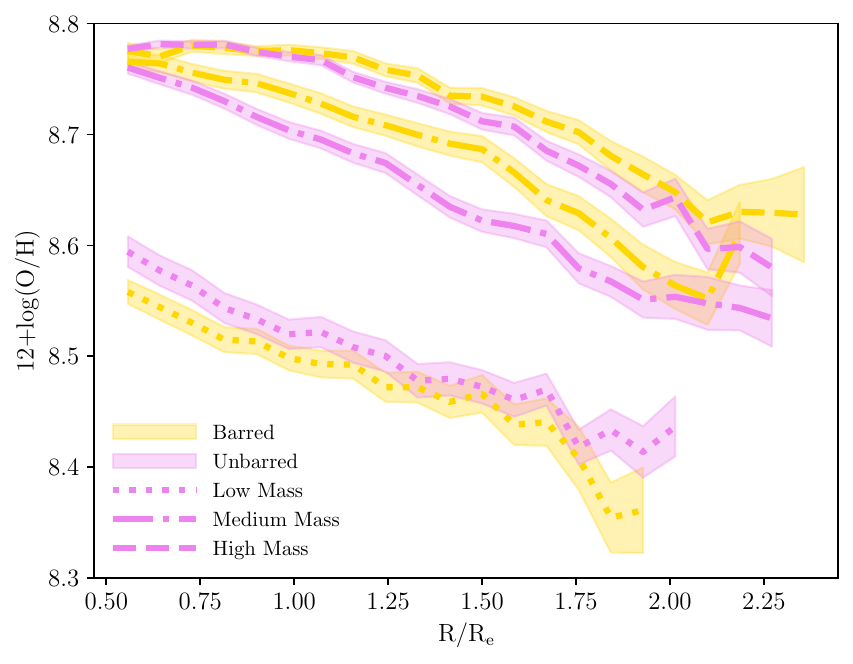}
    \caption{O3N2 Radial metallicity trends for barred (yellow) and unbarred (violet) spiral galaxies for three mass bins, where the different line styles indicate the different mass ranges. The legend shows how line-styles correspond to mass range.}
    \label{fig:o3n2 barred unbarred trends}
\end{figure}

We create radial metallicity trends for our samples of (strongly) barred and unbarred spiral galaxy samples (defined in Section \ref{sec:sample selection}). As expected by the MZR, there are higher metallicity values at $R = R_e$ for both barred and unbarred spiral galaxies with increasing mass, but no obvious difference between the two subsamples. We observe a slight increase in gradient values of the fit lines with increasing mass, and find that unbarred spirals have slightly steeper gradients at all masses (see Figure \ref{fig:bar slope int}; the unbarred samples has an average gradient of $-0.026\pm0.001$~dex/kpc, compared to $-0.018\pm0.001$~dex/kpc for barred), but these differences are very small, as is clear in Fig. \ref{fig:o3n2 barred unbarred trends} which shows trends in three galaxy mass bins. The  presence or lack of a bar in spiral galaxies does not appear to have any significant impact on the azimuthally averaged radial metallicity gradients of these galaxies.

\subsection{Unbarred Spirals Binned by Arm Winding Level}
With our final sub-samples, we investigate the impacts of arm winding levels in unbarred spiral galaxies on the radial metallicity trends. We chose unbarred spirals only to remove any potential impact of a bar and focus only on testing the impact of spiral-arm pitch angle. 
\begin{figure*}
    \centering
    \includegraphics[width = 7.1 in]{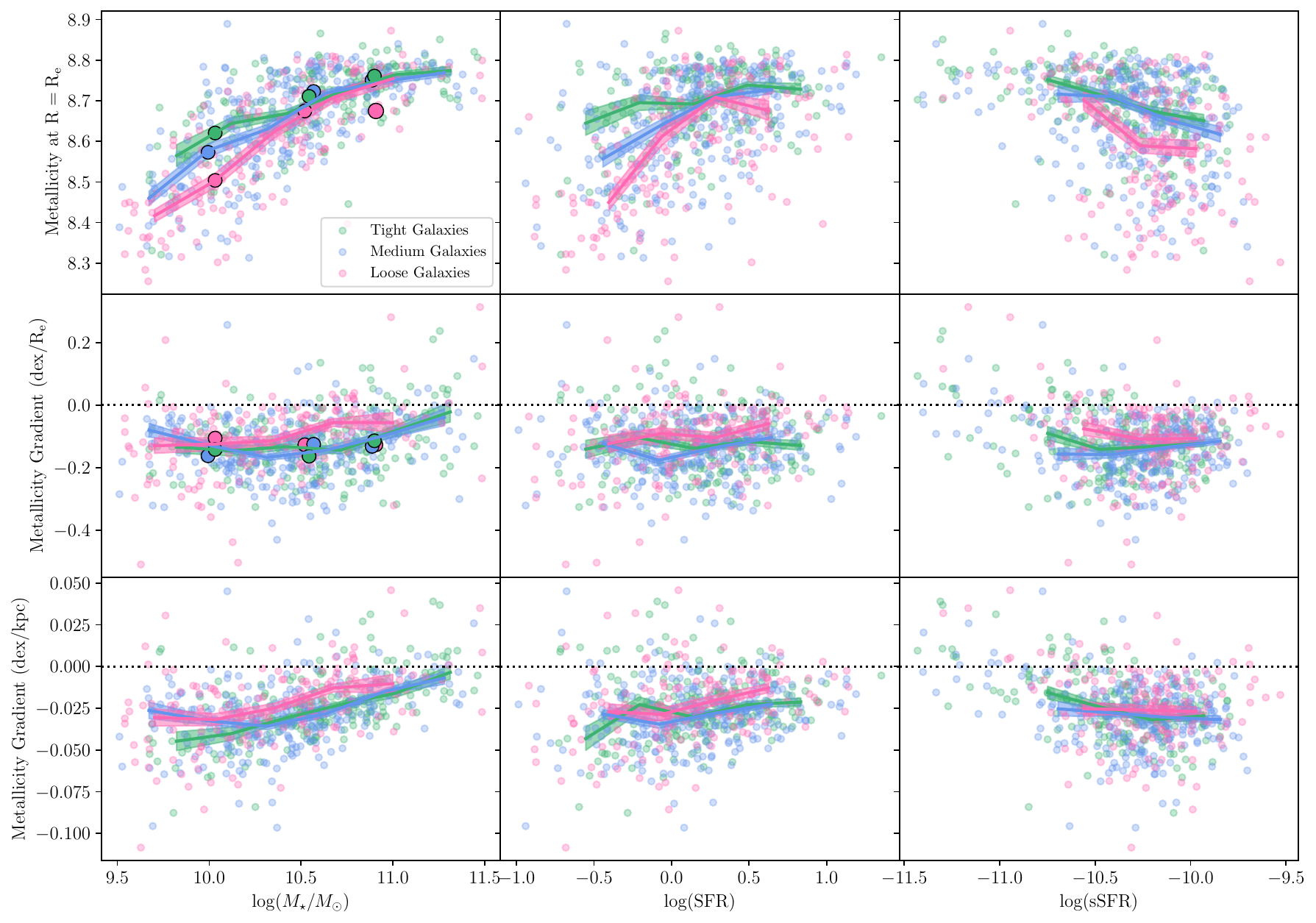}
    \caption{Scatter of metallicity at $R = R_e$ (top row), gradient (middle row: units of dex/$R_e$; bottom row: units of dex/kpc) values from the fit to each galaxy calculated by fitting a binned statistic to the O3N2 based metallicity trends in the sample versus mass (left), SFR (middle), and sSFR (right). Colored lines are binned statistic fit to these points, for each of the three arm winding selections. Values from linear regression fits to trends Fig. \ref{fig:binned_mass_grad} are over-plotted in large circles to show agreement with mass-separated population trends.}
    \label{fig:panel_regression_fits}
\end{figure*} 

We first note that the degree of arm winding correlates with the stellar mass of the spiral galaxies. In our sub-samples, spiral galaxies with more tightly wound arms have higher masses and spiral galaxies with loosely wound arms have lower masses (see the lower left panel in Fig. \ref{fig:mass}). This has been previously noted (e.g. \citealt{Hart2017,Yu2020}), and requires that we consider stellar mass as an important confounding variable in differences with spiral pitch angle. 

We investigate the variation in trends with arm winding first through the linear regressions fit to each individual galaxy in our sample. We plot the metallicity values at $R = R_e$ (top row) and gradients in dex/$R_e$ and dex/kpc (middle and bottom rows) of the regression against the galaxy stellar mass (left), SFR (middle), and sSFR (right) in Fig. \ref{fig:panel_regression_fits}. We use green for galaxies with tightly wound arms, blue for medium, and pink for loose. Simulations suggest that more loosely wound spiral arms should drive more radial relocation and hence galaxies they are found in might have shallower metallicity gradients than in galaxies with the more tightly wound spirals \citep{Danielinprep}. We find only subtle signals of any variation of gas-phase metallicity gradient with arm-winding (in the three samples the average gradients are $-0.024\pm0.002$~dex/kpc for both loose and tightly wound subsets, and  $-0.026\pm0.001$~dex/kpc for medium). We observe that galaxies with loosely wound spirals tend to have slightly shallower gradients, particularly in lower mass galaxies, and where the sSFR is low. However in the highest mass galaxies, and those with higher sSFR, we do not observe any metallicity gradient changes between subsets selected by arm-winding. 

We observe that galaxies with loosely wound arms have notably lower offsets (i.e. metallicity values at $R_e$) than spirals with tighter arm winding, especially in low mass galaxies. This trend is noticeable across all SFR and especially at high sSFRs. 
\begin{figure}
    \centering
    \includegraphics[width = 3.2 in]{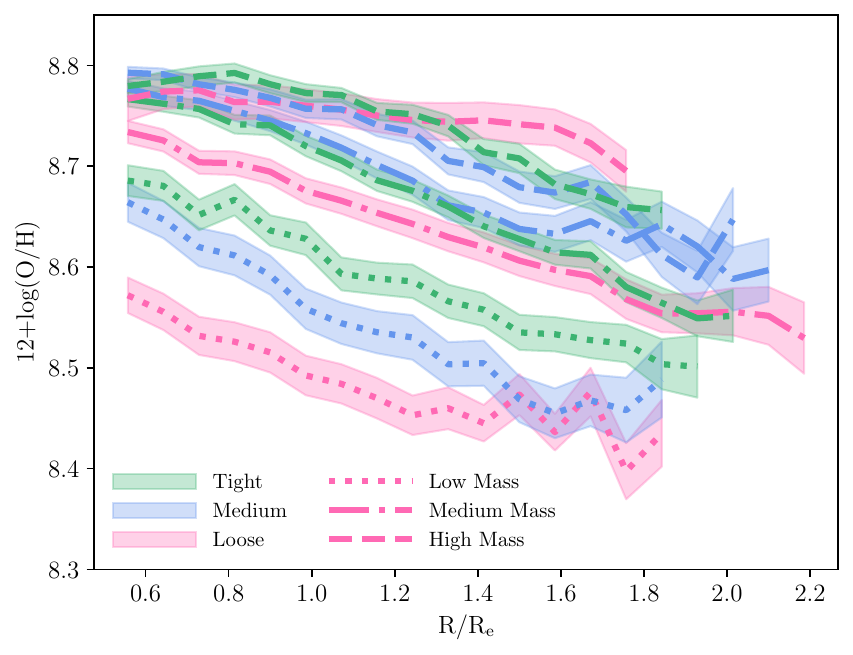}
    \caption{O3N2 binned trends for three mass bin cuts made above for tight, medium, and loose arm winding samples. Colors are same as in Fig. \ref{fig:mass} and \ref{fig:panel_regression_fits}, dotted lines are for $\log (M_\star/M_\odot) < 10.25$, dot-dashed lines for $10.25 < \log (M_\star/M_\odot) < 10.75$, and dashed lines are for $\log (M_\star/M_\odot) > 10.75$ (The legend displays the line style guide in the loose galaxies colorway). }
\label{fig:binned_mass_grad}
\end{figure}

\begin{figure}
    \centering
    \includegraphics[width = 3.2 in]{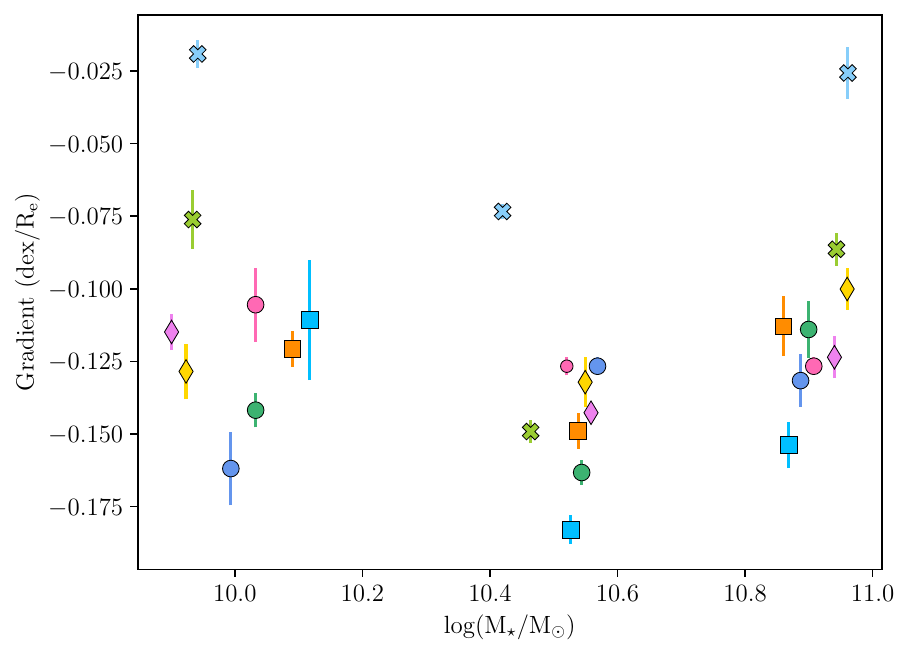}
    \includegraphics[width = 3.2 in]{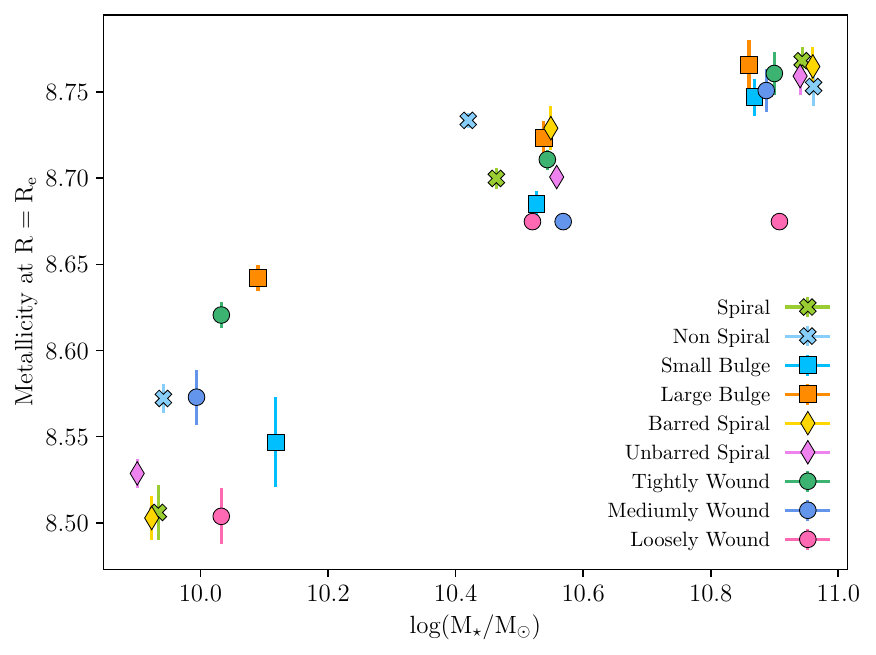}
    \caption{Stellar mass versus gas-phase metallicity gradient (upper; in units of dex/$R_e$) and value at $R/R_e$ (lower panel) from linear regression fits to trends for all discussed sub-samples (shown in Figures \ref{fig:disk v spiral trends mass matched}, \ref{fig:bulge size trends mass matched}, \ref{fig:o3n2 barred unbarred trends} and \ref{fig:binned_mass_grad}). Morphology comparison sub-samples are mass-matched and plotted at the median mass after-matching. Fits are provided in three broad mass bins.}
    \label{fig:trend mass gradient offset}
\end{figure}

Within the mass-matched samples, each mass range in Fig. \ref{fig:binned_mass_grad} shows little variation in the gradients between galaxies with tight, medium, and loosely wound spiral arms, with the exception of higher mass galaxies with loosely wound arms, which appear to have a slightly shallower trend. As in other plots like this, line style indicates one of the three mass bins (dotted lines for the low mass galaxies, dot-dashed lines for the medium mass galaxies, and dashed lines for the high mass sample of galaxies), while the colors show the galaxy arm winding levels. For both the low and medium mass samples, arm winding correlates with metallicity offset, with {\it spiral galaxies with looser wound arms having the lowest metallcities} at fixed mass. 

\section{Summary of Results and Discussion} 
\label{sec:discussion}
Previous work on radial metallicity trends of galaxies, using gas-phase metallicity from IFU surveys, has found that as an overall trend that gradients steepen (become more negative) with increasing stellar mass \citep[e.g.][]{Perez-Montero2016,Belfiore2017}. In this work, we confirm these trends using a subset of 2,632 isolated SF galaxies from the complete MaNGA sample \citep{DR17}. We call this our \emph {SF parent sample}. This sample is comprised of isolated, face-on galaxies which contain enough SF emission line regions to obtain gas-phase metallicity trends. The larger sample allows us to note that these average trends are not well fit by linear relations and we characterize both an `inner' and `outer' change in metallicity (see Section \ref{sec:parent}). We find that in lower mass galaxies ($\log(M_\star/M_\odot<10.3$) metallicity changes on average are actually slightly positive in the outer regions, although averaged over the entire galaxy remain negative, and increasingly so with increasing galaxy mass, while in more massive galaxies ($\log(M_\star/M_\odot>10.3$) they are negative in both inner and outer regions but do not get steeper with galaxy mass in this sample.

From the \emph {SF parent sample}, we create various morphology based sub-samples, mostly using Galaxy Zoo morphologies \citep{Willett2013,Hart2016}. Namely: 
\begin{itemize}
    \item {\it Non spiral galaxy sample} -
    ($N = 878$) - a sample of smooth/non-featured star-forming galaxies (e.g. ``blue ellipticals" or S0s) or star-forming galaxies with features but no visible spirals (e.g. irregular galaxies).
    \item {\it Spiral galaxy sample} -
    ($N = 1233$) - a sample of galaxies with clear spiral structure. From this we select: 
    \begin{itemize}
    \item {\it Spirals with Small/Large Bulges} - $N = 674$ and 559 respectively, samples of spirals with visually identified small or large bulges.  
    \item {\it Barred Spirals} - 
    ($N = 369$) - a sample of spiral galaxies with a distinct bar.
    \item {\it Unbarred spirals} -
    ($N = 664$) - a sample of spiral galaxies with no distinct bar. From this we select: 
     \begin{itemize}
    \item Tightly wound spirals - 
    ($N = 207$) - a sample of unbarred spiral galaxies with tightly wound spiral arms.
    \item Medium Wound Spirals -
    ($N = 286$) - a sample of unbarred spiral galaxies with ambiguously wound spiral arms.
    \item Loosely Wound Spirals -
    ($N = 171$) - a sample of unbarred spiral galaxies with loosely wound spiral arms.
    \end{itemize} 
    \end{itemize}
\end{itemize}

In all cases we mass-match the resulting samples to remove any potential difference caused by the mass-driven gas-phase metallicity gradient trends. The results of linear regressions fit to these mass-matched subsamples in three broad mass bins are shown in Figure \ref{fig:trend mass gradient offset}. These are fits to the trends shown in Figures \ref{fig:disk v spiral trends mass matched}, \ref{fig:bulge size trends mass matched}, \ref{fig:o3n2 barred unbarred trends} and \ref{fig:binned_mass_grad}. 

Summarizing our main observations, we find: 
\begin{itemize}
    \item Galaxies without spirals or visible features have more flattened radial metallicity gradients than galaxies with visible spirals (Figs. \ref{fig:spiral no spiral grad offset}, \ref{fig:disk v spiral trends mass matched}). This is driven by lower metallicities in the outskirts of galaxies with visible spiral arms. 
    \item Spiral galaxies with larger bulges have both higher average metallicities and shallower gradients than those with smaller bulges (see Figures \ref{fig:bulge size slope int}, \ref{fig:bulge size trends mass matched}).
    \item There is no significant difference in the variation of the metallicity gradient at fixed mass for spirals with and without bars (see Fig. \ref{fig:o3n2 barred unbarred trends}). 
    \item We find no significant differences of metallicity radial trends with the tightness of spiral arm winding (see Figs. \ref{fig:panel_regression_fits}, \ref{fig:binned_mass_grad}) with the exception of a reduction in average metallicity values (but no change in radial gradient) in the loosely wound spirals compared to tighter wound spirals, especially notable in the lowest mass galaxy subset. 
\end{itemize}

\subsection{Caveats}
There are a wide range of strong-line gas-phase metallicity calibrations to choose from, outlined in detail in \citet{Kewley2019,Scudder2021}. We ran our analysis using three different measures (O3N2, N2 and $R_{23}$). All three of the metallicity measures that we use are subject to calibration issues caused by diffuse ionized gas (DIG), especially when looking at star forming galaxies. \citet{Zhang2017} investigates the impacts of DIG on SDSS-IV MaNGA star forming galaxies and find that DIG has substantial impacts mostly at higher redshifts. Most notably for N2 based metallicities, there is a possibility of overestimates and systemic flattening of gradients, especially at high redshifts due to the presence of DIG, $R_{23}$ measures are systemically biased, and O3N2 measures show gradient dispersion that can average out for large samples. However, our samples are strictly within a low redshift regime as per the MaNGA sample selection. An additional challenge regarding the use of the $R_{23}$ measure is that this calibrator is double valued, however it is relatively insensitive to the ionization parameter at $12 + \rm log(O/H) > 8.6$, which is in line with values we find using this calibrator. Additionally, \citet{PP2004} finds that N2 and $R_{23}$ are able to calculate metallicity values within the same level of accuracy. We check the results using the three diverse measures.  As we find qualitatively similar results in all three, so we focus on results from just one, picking O3N2. 

 The MaNGA data has a typical physical resolution of 1-2 kpc. This resolution will artificially flatten all observed gradients (although \citet{SanchezMenguiano2016} find no morphological dependence of the flattening of metallicity gradients in the higher physical resolution CALIFA data). 
 
 We are only able to trace gas phase metallicities where emission lines are visible, and distinguishable as being from ionization due to star formation. This means we remove data from $R<0.5 R_e$ and only trace out to $R\sim1.5 R_e$ in most galaxies (the MaNGA primary sample has coverage only to $R\sim1.5 R_e$, see \citealt{Wake2017}). This lower limit on radial range may mean we exclude data from the parts of galaxies that would be most impacted by bar related systems and effects.

 Part of our analysis uses linear fits to the gas-phase metallicity trends as a first order parameterization, then considers how gradients of these fits change from galaxy to galaxy. As is clear in many of the trends we show (e.g. for a single galaxy in Figure \ref{fig:exampletrends}, or in subsets by mass in Figure \ref{fig:SF parent sample trends}) many of the trends show non-linear shapes - like the flattening that starts between $1-1.25R_e$ in Figure \ref{fig:exampletrends}). In future work more complex modeling fitting, or even non-parametric characterization of the trend shapes may reveal other interesting characteristics of these radial gas-phase metallicity trends.

\subsection{Discussion}
Metallicity gradients in galaxies are an important constraint for our understanding of both galaxy formation and evolution. They provide a tracer of the relative importance of inflows, outflows and SFR as a function of radius in a galaxy, but also as a potential signature of radial redistribution or relocation (i.e. changes in average stellar orbit radial size over a star's lifetime). Since disk galaxies are believed to form inside out \citep[e.g.][]{Larson1976,Matteucci1989, Boissier1999, Pezzulli2016, Nedkova2024}, without radial relocation processes local metal enrichment due to in-situ star formation during inside-out disk formation will set up negative metallicity gradients \citep{Franchetto2021}. Radial redistribution/mixing is expected to flatten these metallicity gradients (both stellar and gas phase) by bringing more metal enriched stars or gas outwards (to then seed the ISM with metals), or more pristine gas, or metal poor stars inwards. 

Radial redistribution can be driven by any process which causes changes in angular momentum. One important source of angular momentum transfer is via inhomogeneities in the disk's mass distribution caused by non-axisymmetric morphological features, e.g. spiral patterns and bars. Most mechanisms for radial redistribution \citep{SpitzerSchwarzchild1953,BarbanisWoltjer1967,Lynden-BellKalnajs1972,CarlbergSellwood1985} induce changes in both orbital size and orbital eccentricity of stars leading to a radial diffusion of stars across the disk. On the other hand, transient spiral patterns can drive large changes in orbital size on relatively short timescales \citep{SellwoodBinney2002,DanielWyse2015} through a process sometimes called ``churning" \cite{SchonrichBinney2009} or ``cold torquing" \citep{daniel2019}. Expectations around the efficacy of cold torquing are fundamentally based on the notion that spiral arms have pattern speeds ($\Omega_p={\rm const}$) that are radially independent, such as in the case for density waves.   In this case, the degree of cold torquing is predicted to positively scale with the pitch angle of spiral arms, \citep{Danielinprep}. However, for co-rotating spirals that have pattern speeds that follow the rotation curve ($\Omega_p(R)=\Omega(R)$), such as winding material arms, the degree of orbital reorganization increases as the arms wind up so that the impact of cold torquing is maximized at the lowest pitch angles. The same internal structures that generate radial mixing of stars can also drive radial gas flows (e.g. bars; \citealt{Athanassoula1992,Fragkoudi2016} or spirals; \citealt{Orr2022}).

 Since increased radial relocation of stars and gas should flatten any metallicity gradient these results suggest we should expect galaxies with visible spirals, particularly loosely wound ones, and/or bars to show flatter radial gas-phase metallicity gradients than those without. However we observe none of the simplest expected trends in the gas of these disks. Galaxies with visible spirals are found to present steeper gas-phase metallicity gradients (driven by lower values in the outskirts) and we find no detectable differences in gradient for barred/unbarred and with spiral arm winding). However, while interesting and presenting new puzzles to explain the impact of morphology on galaxy evolution, it is possible that the absence of the simplest expected morphological dependence of the metallicity gradients is not proof of an absence of morphologically driven radial relocation, for a variety of reasons: 

 \begin{enumerate}
    \item We are assuming that all radial relocation flattens metallicity gradients. There is also an implicit assumption that galaxies with the same mass, but different morphologies, should have started from similar initial conditions (in terms of metallicity gradient). In FIRE-2 Milky Way-mass simulated galaxies, \citet{Bellardini2021} find strong negative gradients at all times, with a steeping over time as the galaxies become more rotationally supported. This may hint we are catching galaxies with different morphology at different dynamical life stages. \citet{Iles2023} look at the impacts of bar morphology on stellar migration and find for simulated isolated and tidally impacted galaxies with an initial constant metallicity there are higher metallicities earlier on in the inner galaxy at the time of formation of a bar. 
    
  \item We are considering gas phase metallicity gradients in this work. The ISM is enriched by the previous generation of stellar deaths, hence there is a time delay between ISM enrichment and any large scale relocation of stars. Gas flows both in and out of a galaxy, and gas redistribution which may differ from stellar redistribution may also complicate the interpretation as we discuss more below. Stellar metallicity gradients may be more sensitive to morphology. This question will be investigated in future work. For examples of work looking at stellar metallicity gradients with MaNGA data we mention \citet{Parikh2021,Neumann2021,Lu2023} which consider the impact of global galaxy morphology, but not non-axisymmetric features. 

  \item Regarding our lack of signal from spiral pitch angle, our sample of spiral galaxies only display pitch angles of $14-26^\circ$, much smaller than the range in \citet{Danielinprep}, which ran simulations with set pitch angles of $10^\circ, 20^\circ, 30^\circ, \rm{and}$ $40^\circ$ and for winding arms in a similar range. There is significant scatter in measurements of pitch angle in any realistic galaxy, using any method, and we are unable to determine if an observed galaxy has a fixed spiral pattern speed or winding arms.   Thus, the null change in flattening trends motivated by arm winding levels in the unbarred spiral sample may be a result of the scale of the differences in pitch angles, and the mixing of types of spirals. These samples do not contain galaxies with as large pitch angles as can be set in the simulations in \citet{Danielinprep}. A larger range of observed pitch angles may be necessary to motivate a difference in the gradient of radial trends based on spiral arm pitch angle.

 \item We limit our sample to isolated galaxies (see Section \ref{sec:sample selection}) to mitigate the impact of environmental processes, however galaxies may still have different interaction histories.  Radial flows have been shown to result from galactic tidal interactions \citep[e.g.][]{DiMatteo2007}.  These environmental processes (tidal interactions and mergers) can drive inflow of gas in galaxies \citep[e.g.][]{Montuori2010,Perez2011,Bustamante2018} and there is some observational evidence for this \citep[e.g.][based on measurements of central metallicity]{Michel-Dansac2008}, or \citet{Buck2023} who find a steepening of metallicity gradients at early times due to merger impacts.. \citet{Rupke2010} measured significantly flatter gas phase oxygen abundance gradients in a sample of 16 spiral galaxies experiencing interactions compared to an isolated control sample, however \citet{Schaefer2019} used MaNGA data to show that while satellite galaxies showed a significant increase in central metallicity relative to centrals at the same mass, there were no clear differences in gradients. MaNGA galaxies have also been seen to display a mass-size relation where in a given mass bin, smaller galaxies display flatter gradients \citep{Boardman2021}, which may be driven by the different environments typical of extended (low density) vs. compact (higher density) galaxies \citep{Boardman2023}. Alongside the process of tidally driven redistribution of stars and gas, a galaxy's environment will have an impact on its history of inflow of pristine gas from large scale structure. This cosmological inflow may also impact metallicity gradients, by bringing lower metallicity gas into a galaxy, and has previously been linked to declining metallicity gradients in gas rich galaxies \citep[e.g.][]{Moran2012, Lutz2021}. 

 \item We find no evidence for changes in the azimuthally averaged gas-phase metallicity gradient between barred and unbarred galaxies, but the potential for bars to change metallicity gradients may primarily be along their major axis \cite[e.g.,][]{Filion23}. For example \citet{Fraser-McKelvie2019} found flatter stellar metallicity gradients within bars and within the disc region at the same radius. Bars may not all have the same impact on metallicity gradients \cite[e.g. as noted by][]{Chen2023}. More observational work will be needed to conclude how bars impact gas-phase metallicity gradients.
 \end{enumerate}

  The lower metallicity observed for looser wound arms and higher metallicity for tight arms might suggest that looser wound arms are preferentially found in galaxies which have recently accreted a lot of pristine gas, or could be a sign of radial relocation of gas moving pristine gas inwards more easily in these spirals. If there is more gas accretion in galaxies with more loosely wound arms they might have different cold gas contents. Testing how gas content varies by spiral arm winding would be interesting future work. This observation could be interpreted as evidence that spiral arms wind up over time, behaving as material arms rather than density waves, because older galaxies are more enriched and have higher metallicities, even in narrow mass bins, however it should be noted that the timescales for winding are much shorter than for enrichment.
  
\subsubsection{Comparison with Previous Work}
In the literature, there is some evidence connecting galaxy morphology with systematic variations in the radial trends of gas phase metallicity. These results to date have largely focused on global galaxy morphology (T-type, or bulge size); our study is the largest by far to consider the presence of a bar and the only one we are aware of considering the presence and pitch angle of spiral arm features. We provide here a comparison to a selection of previous results. We also point the reader to the introduction and discussion of \citet{Zurita2021} who provide a comprehensive review of earlier work looking for observational evidence on how bars impact gas-phase metallicity gradients in small samples of galaxies.

Using CALIFA data, \citet{Perez-Montero2016} found that later-type low-mass objects (i.e. late-type spirals and/or irregular galaxies) tend to have flatter slopes (dex/$R_e$) of their metallicity trends; they also noted no dependence on the presence of a bar. This latter observation agrees with our finding, however we find that smaller bulges correlate with steeper slopes. 

Also using CALIFA, \citet{Barrera-Ballesteros2023} investigated the radial profiles of various physical properties in MaNGA data, including how gas phase metallicity trends varied with some measures of morphology. They concluded that only early-type galaxies (early-type star forming galaxies) show significant dependence of gas-phase metallicity gradient on stellar mass -- in late-types, mass had little impact on radial trends. 

In contrast with our work, \citet{SanchezMenguiano2016} found that, in a sample of 112 CALIFA galaxies, gradients were independent of the details of the morphology of the disk galaxy in their sample, and \citet{Sanchez-Menguiano2018} found similar results for another 102 spiral galaxies with MUSE data. 

The work in \citet{Boardman2021} using MaNGA data mostly focused on how a galaxy's radial extent impacted metallicity gradients, but also noted that since late-type galaxies tend to be more radially extended at fixed mass, this could be driving the changes they saw in gradients with galaxy T-type. 

In a pair of papers, \citet{Zurita2020,Zurita2021} use a sample of 51 very nearby galaxies to investigate the effects of bars and spirals on gas-phase metallicity abundances measured in HII regions in those galaxies (the profile data are presented in \citealt{Zurita2020}). \citet{Zurita2021} report a clear flattening of gradients among the lowest luminosity barred galaxies in their sample (relative to unbarred galaxies of the same luminosity), while noting that at higher luminosities there was no difference in gradients. They also note that this trend correlates with spiral type of the galaxies with grand-design spirals tending to be both high luminosity and all have shallow gradients (regardless of the presence of a bar), while the presence of a bar in flocculent spirals (a type of spiral typically found at low luminosity) shallows the gradient. This conclusion appears significantly different from ours, as we observe no significant flattening of the gradients between unbarred and strongly barred galaxies at any mass range (e.g. see Figure \ref{fig:bar slope int},\ref{fig:o3n2 barred unbarred trends}). We also see a steepening of gradients in the subset with obvious spirals (likely containing any grand-design spirals) over the sample of SF galaxies without visible spirals (which likely contains a mix of ``smooth" or early-type SF galaxies, irregular galaxies, and discs with indistinguishable spirals, which might contain some of the flocculent galaxies; see Figure \ref{fig:spiral no spiral grad offset}, \ref{fig:disk v spiral trends mass matched}). There are many differences in both the data and the samples between our study and \citet{Zurita2021} might explain these different conclusions. One is the size of the samples: our spiral sample contains 1233 galaxies, of which 369 are strongly barred spirals, and 664 unbarred (we remove ambiguous/weakly barred spirals completely in considering the impact of the bar), compared to our non-spiral SF sample of 878 galaxies, while \citet{Zurita2021} consider gradients in 51 galaxies in total; 22 strongly barred, 9 weakly barred, and 20 unbarred galaxies. Our methods for both bar and spiral identification also differ considerably. Another major difference which could have an impact is how the spatial resolution and typical radial extent of the metallicity data differ: the data tables provided with \citet{Zurita2020} suggest their data extend to larger radii than the MaNGA data allow (beyond 1.5-2$R_e$), and they measure metallicities in individual HII regions, while MaNGA based gas-phase metallicities are averages over spaxels of 1-2kpc in size. This comparison however suggests that observational tests of these ideas are not yet reaching consensus, and further study will be needed to settle the debate.

\subsection{The Impact of Spiral Features}
We find two intriguing changes in gas-phase metallicity trends linked to the presence or properties of spiral arms: (1) We find that galaxies with visible spirals show steeper radial metallicity gradients, driven by lower metallicities in their outskirts at fixed mass;  (2) We find notably lower average metallicities in low mass spirals with more loosely wound arms relative to more tightly wound at fixed mass. 

 The steeper gradients in galaxies with visible spirals, since it is driven by lower values in the outskirts than the non-spirals, suggests these might have recently accreted more pristine gas. It may point to the link between spiral structure and gas content - the presence of gas is often credited with making spirals more easily visible (e.g. see \citealt{Sellwood2022} for a review). 

 A possible explanation for looser wound spirals having lower average metallicities is that these represent younger spiral galaxies (hence with lower levels of enrichment), which is suggestive of support for the idea that spiral arms wind up over time. Another possibility is that something about the process which generates looser spiral arms makes it more likely to happen in galaxies which have, to date, formed fewer stars, or accreted more recently from the inter-galactic medium (IGM) - hence have lower metallicities - while tighter spiral arms form in galaxies with more typical star formation/accretion histories. 
 
\section{Conclusions}
In this paper we create samples of MaNGA galaxies separate by morphological classifications from Galaxy Zoo and calculate radial gas-phase metallicity trends. The motivation for this work is that morphology might be expected to impact observed gas phase metallicity gradients at fixed mass if different morphological features drive different amounts of radial relocation of stars or gas in a galaxy.

We find consistent results across all three metallicity calibrators. In our parent sample of 2632 isolated star forming galaxies, we confirm previously known trend that gas phase metallicity gradients steepen overall with stellar mass, although note that these trends are far from linear, lower mass galaxies actually show positive metallicity changes in the outer regions and the trend appears to stop or even reverse for masses above $\log M_\star/M_odot = 10.3$. For each sub-classification of morphology we mass-match sub-samples then compare population-wide radial metallicity trends at fixed mass. We find (1) the presence of spiral structure correlates with steeper gas phase metallicity gradients; (2) spiral galaxies with larger bulges have higher gas-phase metallicities and shallower gradients; (3) there is no observable difference with azimuthally averaged radial gradients between barred and unbarred spirals and (4) there is no significant difference in gradient between tight and loosely wound spirals, but looser wound spirals have lower average gas-phase metallicity values at fixed mass. 

This work provides a detailed look at how internal morphologies (bulges, spiral arms and/or bars) correlate with gas-phase metallicity trends in isolated SF galaxies in the MaNGA sample. Although we don't observe expected trends based on the simplest toy models, we find some interesting results showing how the presence and properties of spiral arms in particular change gas-phase metallicity radial trends. Future work investigating stellar metallicity gradients, and/or using new Galaxy Zoo classifications based on the deeper Dark Energy Spectroscopic Instrument (DESI) legacy imaging surveys \citep{Walmsley2023} may help to reveal more details of these interesting results. 

\begin{acknowledgments}
This publication made use of SDSS-IV data. Funding for the Sloan Digital Sky Survey IV has been provided by the Alfred P. Sloan Foundation, the U.S. Department of Energy Office of 
Science, and the Participating Institutions. SDSS-IV acknowledges support and resources from the Center for High Performance Computing at the University of Utah. The SDSS website is www.sdss4.org. SDSS-IV is managed by the Astrophysical Research Consortium for the Participating Institutions of the SDSS-IV Collaboration.

This publication uses Galaxy Zoo 2 data generated via the Zooniverse.org platform, development of which is funded by generous support, including a Global Impact Award from Google, and by a grant from the Alfred P. Sloan Foundation. This data in this paper are in part the result of the efforts of more than 200,000 volunteers in the Galaxy Zoo project, without whom none of this work would be possible. Their contributions are individually acknowledged at {\tt http://authors.galaxyzoo.org}.

EW acknowledges funding from the 2022 Bryn Mawr Summer Science Research program. We wish to acknowledge the use of sample mass-matching code written by Brooke D. Simmons. 

\end{acknowledgments}
\vspace{5mm}
\facilities{SDSS} 
\software{{\tt astropy} \citep{astropy2013, astropy2018, astropy2022}, 
 {\tt Marvin} \citep{Cherinka2019}}
\bibliography{references}{}
\bibliographystyle{aasjournal}

\end{document}